# Chemical design and magnetic ordering in thin layers of 2D MOFs

Javier López-Cabrelles,$^{\$1}$ Samuel Mañas-Valero,$^{\$1}$ Iñigo J. Vitórica-Yrezábal,$^2$ Makars Šiškins,$^3$ Martin Lee,$^3$ Peter G. Steeneken,$^{3,4}$ Herre S. J. van der Zant,$^3$ Guillermo Mínguez Espallargas*$^1$ and Eugenio Coronado*$^1$

$^1$Instituto de Ciencia Molecular (ICMol), Universidad de Valencia, c/Catedrático José Beltrán, 2, 46980 Paterna, Spain.

$^2$School of Chemistry, University of Manchester, Manchester, UK.

$^3$Kavli Institute of Nanoscience, Delft University of Technology, Lorentzweg 1, 2628 CJ, Delft, The Netherlands.

$^4$Department of Precision and Microsystems Engineering, Delft University of Technology, Mekelweg 2, 2628 CD, Delft, The Netherlands.

$^\$$Equally contributed



**ABSTRACT:** Through rational chemical design, and thanks to the hybrid nature of metal-organic frameworks (MOFs), it is possible to prepare molecule-based 2D magnetic materials stable at ambient conditions. Here, we illustrate the versatility of this approach by changing both the metallic nodes and the ligands in a family of layered MOFs that allows the tuning of their magnetic properties. Specifically, the reaction of benzimidazole-type ligands with different metal centres ($M^{II}$ = Fe, Co, Mn, Zn) in a solvent-free synthesis produces a family of crystalline materials, denoted as MUV-1(M), which order antiferromagnetically with critical temperatures that depend on M. Furthermore, the incorporation of additional substituents in the ligand results in a novel system, denoted as MUV-8, formed by covalently bound magnetic double-layers interconnected by van der Waals interactions, a topology that is very rare in the field of 2D materials and unprecedented for 2D magnets. These layered materials are robust enough to be mechanically exfoliated down to a few layers with large lateral dimensions. Finally, the robustness and crystallinity of these layered MOFs allow the fabrication of nanomechanical resonators that can be used to detect —through laser interferometry— the magnetic order in thin layers of these 2D molecule-based antiferromagnets.


## INTRODUCTION

The isolation of atomically thin —single or few-layer— crystals has given rise to the emergence of the so-called two-dimensional (2D) materials. These low-dimensional materials have shown a wide range of electronic and magnetic properties (from insulators to superconductors; from ferromagnets to antiferromagnets and quantum spin liquids) that can be affected by the dimensionality.[1–5] Most of these materials derive from well-known layered inorganic materials,[6] formed by covalently bound layers interconnected by weak van der Waals interactions, allowing the isolation of monolayers by exfoliation. A hot topic in this area deals with the isolation of monolayers of magnets with the aim of studying magnetism in the 2D limit and to integrate these layers in spintronic devices.[7,8] The exploration of such experimental possibilities is hindered by the chemical instability in open air of the chosen inorganic compounds (mainly based on layered metal halides), thus limiting their manipulation and applicability. In fact, the first report on the magnetism of $CrI_3$ down to the monolayer limit was published in 2017[9] even though the discovery of its bulk properties has been known since 1959.[10]

Coordination chemistry can also offer a source of layered metal-organic materials which, in contrast to the inorganic analogs, are much more versatile from the point of view of the chemical design, chemical stability and ease of manipulation in open air. However, the isolation of monolayers of these molecule-based magnets has been achieved in very few cases only.[11] This may be due to the hybrid nature of the materials, which in most cases are formed by charged layers interleaved by counter-ions. Hence, these layers are held together by electrostatic interactions and not by van der Waals interactions. This often results in fragile crystals of small sizes, which are very difficult to exfoliate using a micromechanical Scotch tape procedure. In fact, the first successful mechanical exfoliation of a layered magnet of this kind was reported by us in 2015 in the coordination polymer $[Fe^{III}(acac_2\text{-trien})][Mn^{II}Cr^{III}(anilate)_3]\cdot(CH_3CN)_2$, where anilate refers to dichloro- and dibromo-substituted anilate ligand.[12] These compounds in bulk behave as ferrimagnets with a transition temperature, $T_c$ = 11 K. The structure of the magnetic layers consists of a hexagonal 2D anionic network formed by $Mn^{2+}$ and $Cr^{3+}$ ions linked through anilate ligands. In this case, and thanks to the large size of the hexagonal pores, the counter-ions $[Fe^{III}(acac_2\text{-trien})]^+$ were inserted inside the magnetic framework instead of being in the interlamellar space. This feature facilitated the micromechanical exfoliation, leading to the isolation of magnetic

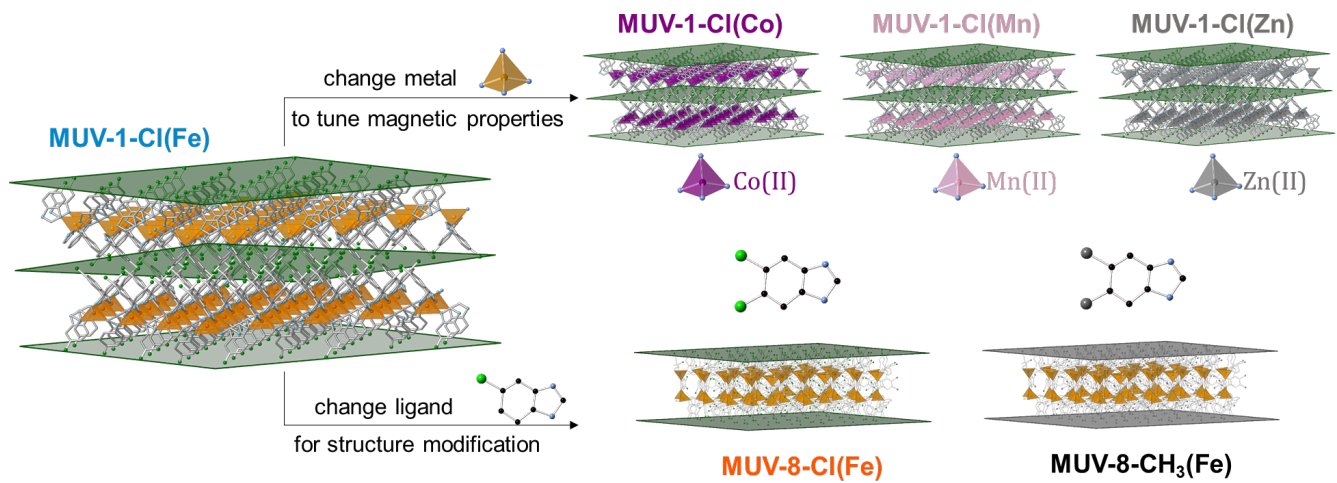

**Figure 1**. Chemical versatility of the MUV-1 system: it is possible to modify the magnetic properties by changing the metallic cation (upper part) and to induce structural changes by adding a second substituent to the benzimidazole ligand (bottom part), while keeping the layered morphology needed for a two-dimensional material.

layers with thicknesses down to 1.5 nm and lateral sizes on the order of hundreds of nanometers.

More recently, this result has been drastically improved by designing coordination compounds formed by neutral layers.[13] To reach this goal we have exploited the chemical versatility provided by imidazolate-type ligands in the design of metal organic frameworks (MOFs) of various dimensionalities and properties.[14–16] Thus, a isoreticular series of layered magnetic MOFs, composed by benzimidazole derivatives and Fe(II) centres (**MUV-1-X(Fe)**, with X = H, Cl, Br, CH$_3$ or NH$_2$) and behaving as spin-canted antiferromagnets with ordering Neel temperatures, $T_N$, of ~20 K, has been synthesized by a solvent-free method.[13] In this series, monolayers have been isolated through micromechanical exfoliation, reaching high-quality crystalline flakes of micron size. On the other hand, we have also shown that these materials can be exfoliated in large amounts by using a liquid exfoliation method.[17]

Interestingly, the chemical versatility of this molecular approach has allowed us to functionalize the magnetic layer at will by changing the substituent X in the benzimidazole, while keeping its magnetic properties unchanged.[13] Herein, we further exploit this versatility either by changing the metallic nodes, while maintaining the crystal structure, or by inserting a second substituent in the ligand, while retaining the layered morphology of the material (Figure 1). The first possibility provides the opportunity to tune the magnetic properties of the layers, while the second results in the isolation of covalently bound magnetic double-layers interconnected by van der Waals interactions, a topology that is very rare in the field of 2D materials and unprecedented for 2D magnets. In the second part of this work, these van der Waals antiferromagnets are mechanically exfoliated down to the atomically-thin layers and we show that their magnetic ordering can be probed mechanically by nanomechanical resonators made of ultrathin membranes of these insulating materials.

**Changing metallic nodes**. The existence of different metal sources compatible with chemical vapor deposition and solvent-free methods permits the modification of the reaction previously described in ref.[13], where benzimidazole derivatives (HbimX, X = Cl, Br, CH$_3$, H, NH$_2$) and ferrocene were employed, using different metal cyclopendienyls and other metal precursors (see details in Section S1). Obtaining isostructural materials changing the metal node is a challenge that leads to the modification of the properties of the materials; in the present case, the magnetic properties (Figure 1). Biscyclopentadienyl cobalt (II), bis(tetramethylcyclopentadienyl) manganese (II), bis(2,2,6,6-tetramethyl-3,5-heptanedionato) zinc (II) and ZnO have been used as metal precursors in a solvent-free method to synthetize isostructural compounds of **MUV-1-Cl** and **MUV-1-H**. The crystals obtained in the synthesis were characterized by single crystal X-ray diffraction and/or powder X-ray diffraction (see Section S2 and S3), obtaining three isostructural compounds of **MUV-1-Cl** (Co, Mn and Zn) and three of **MUV-1-H** (Co, Mn and Zn). The metal cation is located in the inner part of the layers in a distorted tetrahedral environment, connected by benzimidazole bridges, allowing magnetic exchange between the metal centers. The size of the crystals is smaller for the cobalt, manganese and zinc compounds than for the previously reported iron analogue, but all of these materials keep the layered morphology (Figure 2 and S1-S3). These changes in the metallic nodes introduce changes in both the single-ion magnetic anisotropy and the intralayer exchange interactions that permit the tunability of the magnetic properties in these materials (see Magnetic properties section).

RESULTS AND DISCUSSION

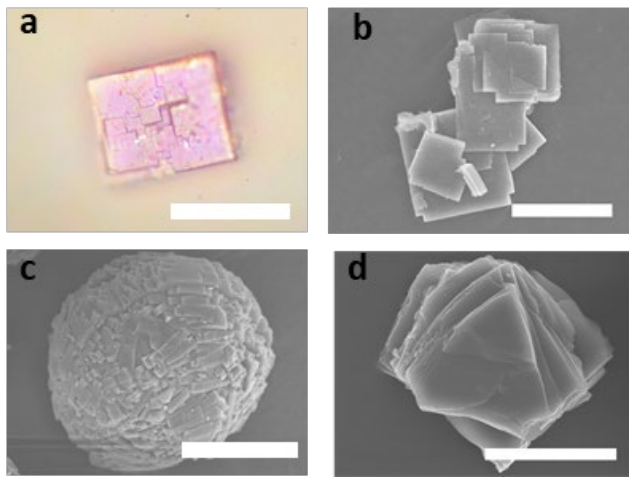

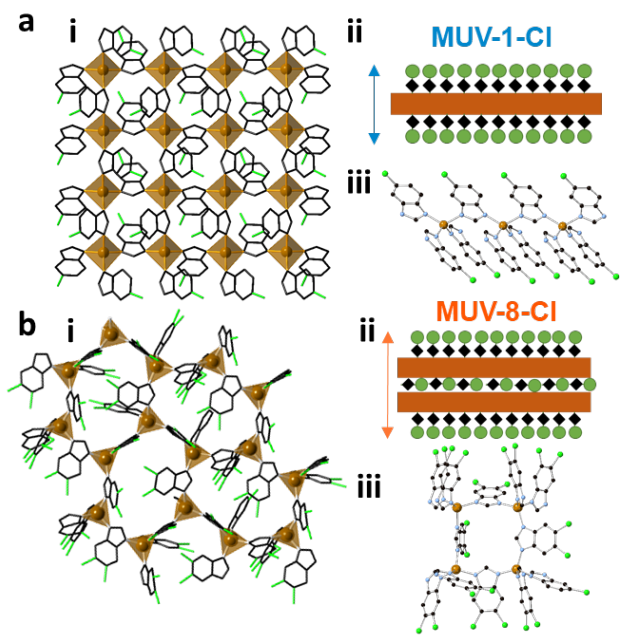

**Figure 2**. (a) Optical images for a crystal of **MUV-1-Cl(Co)**. (b,c) SEM images of **MUV-1-Cl(Co)** and (d) **MUV-8-Cl(Fe)** showing the layered morphology. Note that **MUV-1-Cl(Co)** presents spontaneous crystal aggregation with spherical uncommon morphologies. Scale bars are 10, 10, 20 and 30 μm, respectively.

**Changing the ligands**. The molecular nature of the coordination polymers not only permits changes in the metal nodes but also in the ligand part, (*i.e.,* on the functional groups). In this case, a second substituent is added in the organic ligand adjacent to the 5 position where the initial substituent is present. Thus, two different ligands have been used, 5,6-dichlorobenzimidazole (HbimCl$_2$) and 5,6-dimethylbenzimidazole (Hbim(CH$_3$)$_2$). Importantly, this change in the ligand does not affect the formation of a layered structure, and their reaction with ferrocene gives rise to layered coordination polymers, the so-called **MUV-8-Cl(Fe)** and **MUV-8-CH$_3$(Fe)** (Figure 2 and S1). However, this second functional group induces a significant structural change resulting in unprecedented double layers of iron (II) centers arranged in distorted hexagons and linked through bridging benzimidazole ligands (Figure 3).

There are seven crystallographically different iron centers in a single double layer of **MUV-8** (Figure S2). These iron centers are connected by three bidentate ligands in the *ab* plane, and a fourth bidentate ligand connecting two iron centers each in its respective layer (in the *c* direction) (Figure 3biii). As a consequence, the resulting double layer in **MUV-8** exhibits a thickness of 1.5 nm, being formed by the sequence "ligand layer/ Fe layer/ ligand layer/ Fe layer/ ligand layer", which is large compared to the thickness of 1 nm in **MUV-1** since that is formed just by one monolayer of Fe centers (sequence of "ligand layer/ Fe layer/ ligand layer"). The iron centers are located in very distorted tetrahedral sites, with Fe–N lengths of 1.8–2.15 Å, and connected by benzimidazolate bridges.

**Figure 3.** Comparison of the crystal structures of **MUV-1-Cl** (a) and **MUV-8-Cl** (b). Structure of a single layer (**i**) viewed along the *c* axis (*ab* plane). Schematic representation of a single layer (**ii**), showing an increase thickness for the "double layer" of irons in the **MUV-8-Cl** (from 1nm to 1.5 nm). Coordination modes of the ligands (**iii**).

These connected layers are superposed in a similar arrangement to an AB stacking, with a small shift between iron centres (Figure S2). The chlorine atoms belonging to the ligands placed in the *ab* plane are pointing toward the surfaces, and are weakly interacting through van der Waals forces like in the **MUV-1-Cl** case, while the bridging ligand connecting the two monolayers is located inside of the double layer. Moreover, in **MUV-8-X**, some ferrocene molecules are located between the layers, interacting weakly with the layers of the compound (Figure S3). These molecules cannot be removed with heat as can be seen by thermogravimetric analysis (Figure S6). The two-dimensional units are composed of covalently bound double layers held together through coordination bonds and interconnected by weak van der Waals interactions, instead of being formed by covalently bound monolayers. This novel structural arrangement opens the way to isolate unprecedented 2D magnetic networks using a coordination chemistry approach.

The layered morphology of the crystals (Figure 2) allows their delamination and the possibility to explore the 2D limit in these compounds. We have focused on the cobalt system, **MUV-1-Cl(Co)** and **MUV-1-H(Co)**, and on **MUV-8-Cl(Fe)** and **MUV-8-CH$_3$(Fe)**. Bulk crystals of these four systems were thinned-down by mechanical exfoliation as previously realized in **MUV-1-Cl**,[13] yielding flakes with well-defined shapes (lateral dimensions of > 1 μm) and different thicknesses (from hundreds of nanometers down to a few nanometers). The obtained flakes are characterized by optical and atomic force (AFM) microscopies (Figure 4a and Section S5) as well as Raman spectroscopy, in order to confirm their integrity and chemical composition. In the case of the novel structures of **MUV-8**, we have achieved thin-layer with a thickness of 6 nm,

which in this case corresponds to 3-4 monolayers. Importantly, we have been able to detect the Raman spectra of these ultra-thin films (Figure 4c).

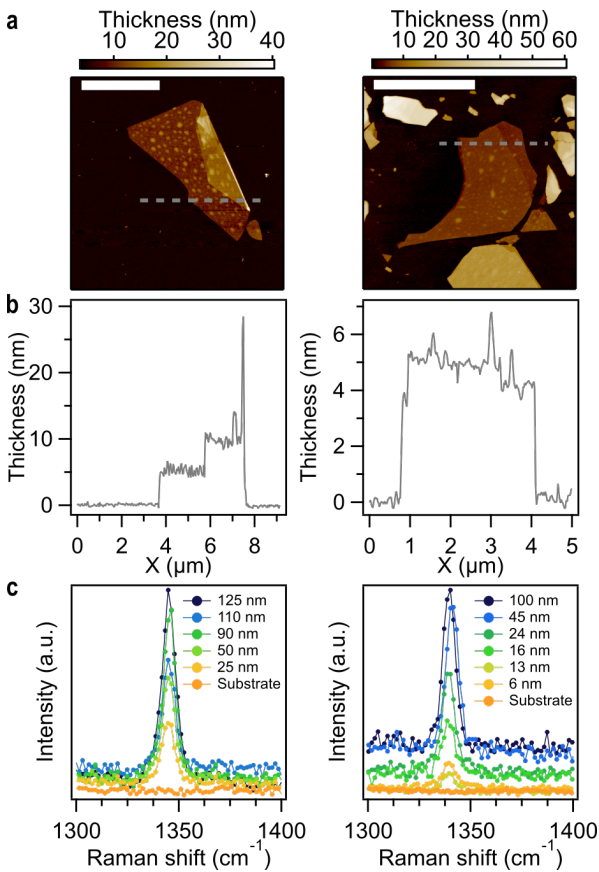

**Figure 4.** (a,b) AFM image and profile of a typical exfoliated **MUV-1-Cl(Co)** (left) and **MUV-8-Cl(Fe)** flake (right). Scale bars are 5 μm. (c) Selected region of the Raman spectra of **MUV-1-Cl(Co)** (top) and **MUV-8-Cl(Fe)** (bottom) flakes of different thicknesses.

**Magnetic properties**. The magnetic behavior of the polycrystalline bulk samples was investigated by SQUID measurements in the range of 2-300 K. Respective data can be found in the supplementary information (see section S4). In Figure 5 we plot the thermal variation of the susceptibility ($\chi$) for the Co(II) and Mn(II) derivatives of **MUV-1**, as well as the Fe(II) derivative of **MUV-8**. In these compounds $\chi$ increases upon cooling down until a maximum is reached. This agrees with an antiferromagnetic coupling between the metal centers, also supported by the continuous decrease in the $\chi T$ product in this region (insets of the Figure). At lower temperatures, a sharp increase in $\chi$ is observed for the Co and Fe samples. This feature, together with the presence of a sharp out-of-phase component of the a.c. susceptibility signal, $\chi''$, at this temperature (Figure 5, right) and magnetic hysteresis loops (Figure S18), are indicative of a transition towards antiferromagnetic order with spin canting at the critical temperature, $T_N$. In contrast, in the Mn derivative, the maximum in $\chi$ is followed by a sharp decrease at lower temperature, which agrees with an antiferromagnetic ordered structure without canting (Figure 5, left). This is corroborated by the absence of a $\chi''$ signal in the vicinity of $T_N$ (Figure 5, right). Such a difference may be related to the fully isotropic nature of the spin in the Mn derivative (with a ground term $^6A_1$), while some magnetic anisotropy is expected for high-spin Co(II) and Fe(II) in tetrahedral sites (described by $^4A_2$ and $^5E$ terms, respectively).

The critical temperatures in these antiferromagnets have been estimated from the temperature where $\chi''$ differs from 0 in the case of systems with spin-canting (**MUV-1(Co)** and **MUV-8(Fe)**). In the case where the canting is absent (**MUV-1(Mn)**), $T_N$ is estimated, according to the Fisher criteria,[18] by considering the maximum in $\partial(\chi T)/\partial T$. In this last system, a $T_N$ value of 14 K is derived, which is consistent with the EPR measurements that show important changes in the signal near this temperature (see Section S4.2). $T_N$ values for all the materials are summarized in Table 1.

Finally, an estimate of the exchange parameters in these 2D antiferromagnetic networks is obtained by fitting the magnetic data to the corresponding theoretical models. To describe the behavior of **MUV-1** derivatives, a model for a quadratic-layer Heisenberg antiferromagnet is used (Lines model,[19] see SI). This model reproduces in particular the rounded maximum observed in $\chi$ in the Mn(II) derivatives (Figure 5b). The exchange parameters for **MUV-8(Fe)** are more difficult to extract since the magnetic network is very complex (see Figure S12). The structure involves two distorted hexagonal lattices, each one formed by 2 different exchange parameters, J and $J_0$; these two layers are coupled by an additional exchange, $J_{inter}$. There is no theoretical model available to describe this network. The closest model available involves a single hexagonal lattice of classical spins isotropically coupled (Curély model,[20] see Section S4). Using this model, an antiferromagnetic coupling J = $J_0$ = –25 cm$^{–1}$ is estimated, which is within the range of those obtained for **MUV-1(Fe)** compounds (see Table 1). As we can see in this table, the exchange coupling in all the compounds is antiferromagnetic. Interestingly, for a given metallic derivative, the magnetic properties remain unaltered, independent of the type of derivatization on the benzimidazole ligand. Still, these properties change upon changing the metal center (from Fe to Mn and Co for **MUV-1**, for example).

**Table 1**. Magnetic properties of the **MUV-1** and **MUV-8** families. J and g parameters are extracted by fitting the magnetic data to Lines[19] and Curély[20] models for quadratic and hexagonal exchange networks, respectively. The form of the exchange Hamiltonian is $\left(\mathcal{H} = -J\sum_{i,j} S_i \cdot S_j\right)$. *MUV-1(Fe)** data have been taken from reference [13].

| MUV | S | $T_N$ (K) | J (cm$^{-1}$) | g |
|---|---|---|---|---|
| MUV-1-Cl(Fe)* | 2 | 20.7 | -22.9 ± 0.4 | 2.00 ± 0.02 |

| Compound | | | | |
|---|---|---|---|---|
| MUV-1-H(Fe)* | 2 | 20.0 | -23.5 ± 0.2 | 1.98 ± 0.02 |
| MUV-1-Br(Fe)* | 2 | 20.0 | -22.8 ± 0.2 | 2.0 ± 0.2 |
| MUV-1-CH$_3$(Fe)* | 2 | 20.1 | -22.6 ± 0.4 | 2.0 ± 0.2 |
| MUV-1-NH$_2$(Fe)* | 2 | 21.2 | -23.3 ± 0.3 | 2.0 ± 0.2 |
| MUV-1-Cl(Co) | 3/2 | 11.6 | -20.2 ± 0.4 | 2.3 ± 0.2 |
| MUV-1-H(Co) | 3/2 | 12.4 | -20.8 ± 0.4 | 2.2 ± 0.2 |
| MUV-1-Cl(Mn) | 5/2 | 14.3 | -10.7 ± 0.2 | 2.2 ± 0.2 |
| MUV-1-H(Mn) | 5/2 | 14.8 | -10.4 ± 0.2 | 2.02 ± 0.10 |
| MUV-8-Cl(Fe) | 2 | 23.2 | -24.6 ± 0.5 | 2.0 ± 0.2 |
| MUV-8-CH$_3$(Fe) | 2 | 23.4 | -25.2 ± 0.5 | 2.1 ± 0.2 |

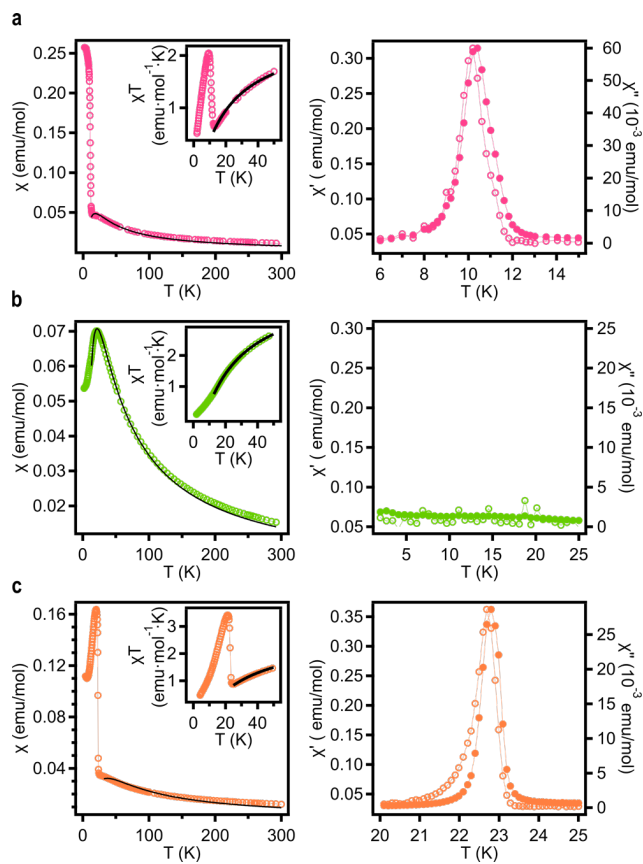

**Figure 5.** d.c. (left column) and a.c. (right column; in-phase and out-of-phase components at 110 Hz are denoted with solid and open circles, respectively) magnetic susceptibilities for different **MUV-1** and **MUV-8** compounds. (a) **MUV-1-Cl(Co)**, (b) **MUV-1-H(Mn)**, (c) **MUV-8-Cl(Fe)**. Solid black lines indicate the fit in the low dimensional regime to the expression for an antiferromagnetic square lattice (**MUV-1**) and the Curély formula for an antiferromagnetic hexagonal layer (**MUV-8**). [19,20]

**Magnetic order by nanomechanical resonators**. The detection of magnetic order in thin-layers of insulating 2D materials is a challenging problem as it is very difficult to sense a so small amount of material by conventional bulk characterization methods (magnetic or specific heat measurements). In fact, in these 2D materials magnetic ordering has been only detected recently in inorganic magnetic materials by performing Magneto-Optical Kerr Effect measurements at the nanoscale (nanoMOKE),[9] transport measurements in van der Waals heterostructures[21–23] or by NV (nitrogen vacancies) magnetometry.[24] Indirect techniques, such as optical measurement of the second harmonic generation, have also been used to characterize thin layers of inorganic antiferromagnets.[25,26] Recently, we have demonstrated that the specific heat can be extracted from measuring the temperature dependent nanomechanical resonance frequency of suspended 2D antiferromagnetic membranes.[27] In particular, it can be shown that the resonance frequency of the fundamental membrane mode, $f_0$, and the quality factor, $Q$, of a nano-mechanical resonator are related to the specific heat, $c_v$, by the following relations: $c_v(T) \propto d(f^2_0(T))/dT$ and $Q^{-1} \propto c_v(T) \times T$. These relations have been used to prove the antiferromagnetic order in the inorganic layers FePS$_3$, MnPS$_3$ and NiPS$_3$.[27] However, the larger fragility of metal-organic materials to lasers has prevented so far to detect magnetic order in thin layers of these molecular magnets. In our previous work, we demonstrated that the **MUV-1** family is robust enough to fabricate mechanical resonators and to measure its mechanical properties through laser interferometry.[13] Taking advantage of this feature —uncommon for a molecular material— we will explore here the possibility to detect magnetic order in these molecular layers using this nano-mechanical technique.

We exfoliate flakes of **MUV-1-Cl(Fe)**, **MUV-8-Cl(Fe)** and **MUV-1-H(Co)** and transfer these on top of circular cavities (diameter d = 5 – 6 µm) etched in a SiO$_2$/Si substrate using deterministic dry viscoelastic stamping to form freestanding nanodrums (see Figure 6a-c).[28] Due to the large flexibility, low estimated Young's modulus and, thus, low bending rigidity of these MOF sheets, the suspended flakes behave as membranes even at relatively large thicknesses for a given range of cavity diameters.[13] The samples are placed in a dry cryostat and cooled down to temperatures of 4 K at a pressure below 10$^{-6}$ mbar. Temperature dependent mechanical properties of the nanodrums are then investigated using laser interferometry[27] from 4 to 50 K in the absence of an external magnetic field (see Methods). Figure 6d-f shows resonances of the fundamental membrane mode at 40 K for **MUV-1-Cl(Fe)**, **MUV-8-Cl(Fe)** and **MUV-1-H(Co)**, respectively (black solid dots).

We fit the measured resonance peaks to a linear harmonic oscillator model (colored solid lines) and obtain $f_0$ and $Q$. Substantially large Q factors ranging from 1000 to 3500 and high resonance frequencies of 32.1 – 40.67 MHz at 40 K indicate a high tension in the membranes due to a build-up of the thermal strain. Since the thermal strain in the membranes is related to the thermal expansion coefficient and thus to the specific heat $c_v$ of the material, we will check if any anomaly is observed in $f_0(T)$, related to a phase change.[21] In Figure 6g-i we plot $f_0$ (colored filled dots) and the corresponding temperature derivative of $f^2_0$ (black filled dots) for all three compounds. The phase transition-related anomaly is well visible as a kink in $f_0(T)$ and is even more pronounced in $d(f^2_0 (T))/dT$, which shows a peak at low temperatures that is associated to $T_N$. As can be seen in Figure 7, the transition temperatures for the thin-layers, extracted from these peaks, are in close agreement with the values determined in table I for the bulk counterparts by a.c. magnetic measurements (**MUV-1-Cl(Fe)**: 19.5 ± 1.0 K, compared to 20.7 K; **MUV-8-Cl(Fe)**: 20.5 ± 1.0 K, compared to 23.2 K; **MUV-1-H(Co)**: 11.0 ± 1.0 K, compared to 12.4 K). The mechanical dissipation $Q^{-1}(T)$ also exhibits a local maximum near $T_N$, as displayed in Figure 6j-l, that can be related to thermoelastic or other more intricate magnetomotive damping mechanism.[27]

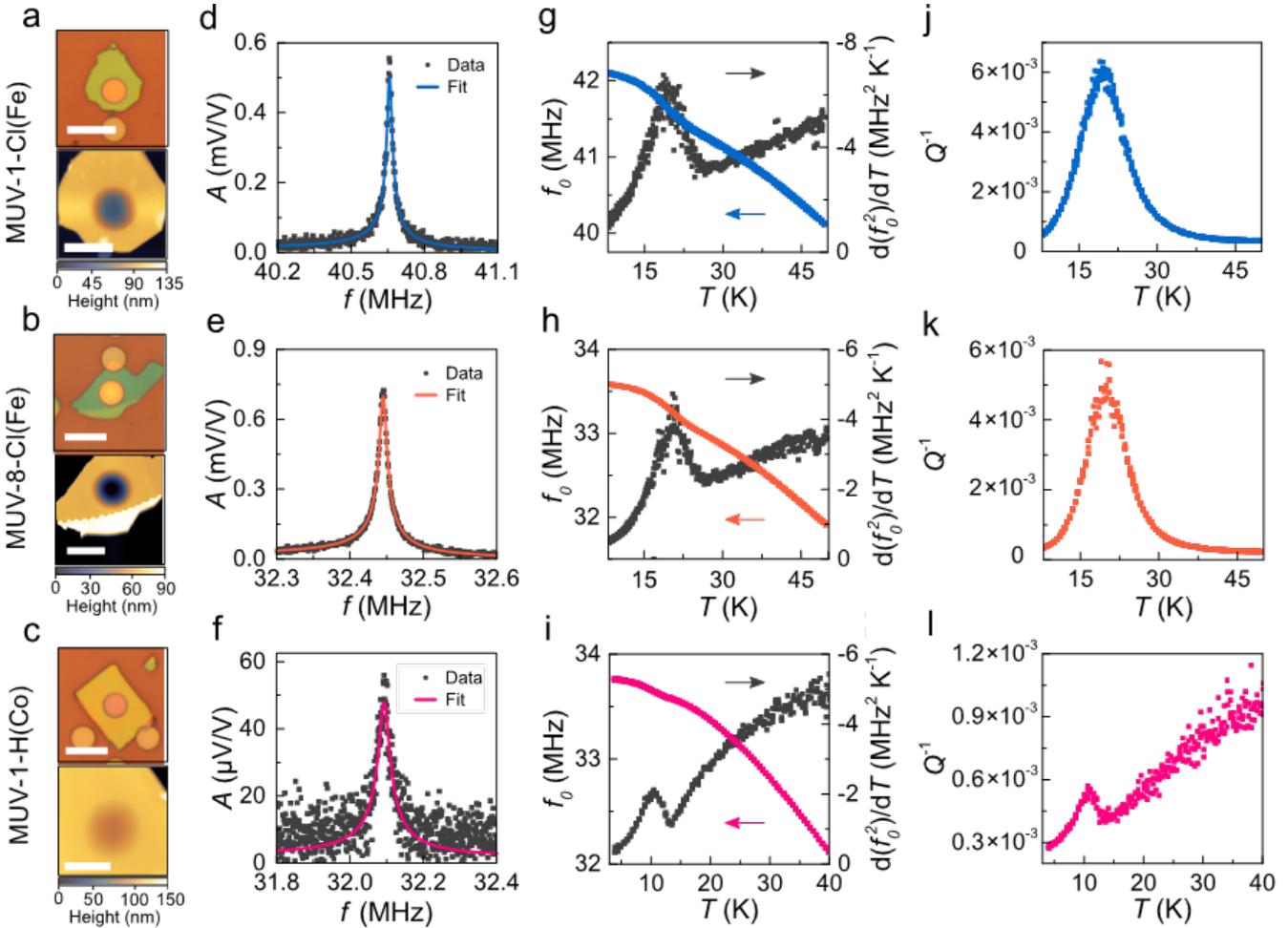

**Figure 6**. Mechanical resonances of thin MOF membranes. (a-c) **MUV-1-Cl(Fe)**, **MUV-8-Cl(Fe)** and **MUV-1-H(Co)** membranes. (a-c) Top panels: Optical images. Bottom panels: Peak-force atomic force microscopy (AFM) images. (a) Scale bars: Top and bottom panel - 10 μm. Membrane thickness $t$ = 87.7±0.7 nm and diameter $d$ = 5 μm. (b) Scale bars: Top panel - 12 μm. Bottom panel - 6 μm. Membrane thickness $t$ = 65.2±1.3 nm and diameter $d$ = 6 μm. (c) Scale bars: Top panel - 12 μm. Bottom panel - 6 μm. Membrane thickness $t$ = 116.4±1.4 nm and diameter $d$ = 6 μm (d-f) Resonance peaks of the fundamental membrane mode at 40 K. Colored lines - linear harmonic oscillator fit, black dots- measured data. (g-i) Colored dots - resonance frequency $f_0$ as a function of temperature, black dots - temperature derivative of $f^2_0$ (T) as a function of temperature. (j-l) Mechanical dissipation $Q^{-1}$ as a function of temperature.

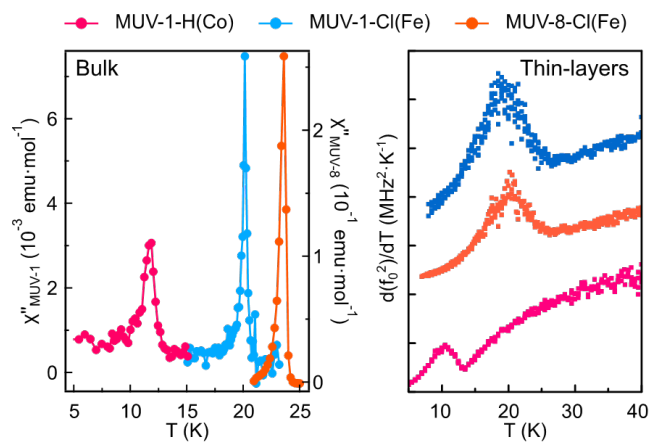

**Figure 7.** Comparison of the out-of-phase a.c. susceptibility signals for bulk systems (left) and the derivative of $f_0^2\,(T)$ as a function of temperature for thin-layers (right).

## CONCLUSIONS

At the intersection between the fields of molecular magnetism and 2D materials, we have exploited the chemical design of layered magnetic coordination polymers based on benzimidazole derivatives with the aim of producing novel 2D molecular-based magnets with novel topologies and tunable magnetic properties. Thus, using a solvent-free synthetic route, we have prepared an isoreticular series of layered materials formed by van der Waals layers with a square 2D magnetic lattice, in which the magnetic properties can be tuned by changing the metallic nodes (from Fe(II) to Co(II) and Mn(II)), while preserving the same crystal structure. In addition, by modifying the organic part (from monosubstituted to disubstituted benzimidazole), it has been possible to change the crystal structure and the magnetic topology from a square lattice to a distorted hexagonal one.

Interestingly, the very weak van der Waals forces between the layers has allowed us to isolate atomically-thin layers of micrometer size flakes using a micromechanical exfoliation method. This result can be highly relevant in the field of 2D materials: First, it can offer an alternative way to that provided by solid-state chemistry for the preparation of robust 2D magnets. In fact, the coordination chemistry approach has afforded the isolation of 2D magnetic materials relatively stable in open air, in sharp contrast to the few reported examples of 2D inorganic magnets. Second, the chemical versatility of the method allows to obtain novel magnetic topologies in two dimensions. The most relevant result in this context has been the isolation of a double layer antiferromagnet exhibiting an unprecedented topology, which is based on the assembly of two hexagonal monolayers through coordination bonds. Third, the employed solvent-free methodology is compatible with chemical vapor deposition techniques, thus allowing the future scalability of these 2D magnetic MOFs.

Nanomechanical resonators made from the MOFs have been used to detect the magnetic order in thin layers of these 2D molecular-based antiferromagnets. Due to the strong coupling between mechanical motion and magnetic order, this technique has provided additional confirmation of the critical temperatures in these thin layer systems. The use of this nanomechanical approach for detecting magnetic order in molecular materials is believed to be limited by their inherent instability to the exposure of the laser beam, which can damage the suspended membranes. Still, the layered MOFs studied have shown to be stable enough to be measured using this technique. In the present case, this has allowed the study of the magnetic order in membranes with layer thickness down to 65 nm, although this thickness is still far from the 2D limit (1-2 nm). More robust and thinner membranes will be prepared in the future taking advantage of the possibility of fabricating van der Waals heterostructures using a deterministic method.[29] For example, by combining ultrathin layers of these MOFs with atomically-thin layers of an inorganic material (h-BN layer, for instance), we expect to overcome this limitation. This will open interesting possibilities in 2D physics. Also, the tuning of $T_N$ with the dimensionality of the system (number of magnetic layers), or with the strain of the membrane induced by an external stimulus (like an electrostatic gate voltage)[30] could in future be realized.

## EXPERIMENTAL SECTION

All reagents were commercially available and were used without further purification.

**Synthesis of MUV-1-X(M$^{II}$).** Metallic source (0.16 mmol) and benzimidazole derivate (0.34 mmol) were combined and sealed under vacuum in a layering tube (4 mm diameter). The mixture was heated at 150 °C for 4 days to obtain crystals suitable for X-ray single-crystal diffraction. The product was allowed to cool to room temperature, and the layering tube was then opened. The unreacted precursors were extracted with acetonitrile and benzene, and the main compound was isolated as crystals (yield 60 %). Phase purity was established by X-ray powder diffraction.

**Synthesis of MUV-8-X(Fe).** Ferrocene (30 mg, 0.16 mmol) and 5,6-dichlorobenzimidazole (64 mg, 0.34 mmol) or 5,6-dimethylbenzimidazole (50 mg, 0.34 mmol) were combined and sealed under vacuum in a layering tube (4 mm diameter). The mixture was heated at 250 °C for 3 days to obtain colourless crystals suitable for X-ray single-crystal diffraction. The product was allowed to cool to room temperature, and the layering tube was then opened. The unreacted precursors were extracted with acetonitrile and benzene, and the main compound was isolated as colourless crystals (yield 60 %). Phase purity was established by X-ray powder diffraction.

**Single crystal X-ray diffraction.** X-ray data for compounds **MUV-1-H(Co)** and **MUV-8-Cl(Fe)** were collected at a temperature of 120 K using a Mo-kα radiation on a Rigaku Supernova diffractometer equipped an Oxford Cryosystems nitrogen flow gas system. X-Ray data for compound **MUV-1-Cl(Mn)** were collected at a temperature of 100 K using a Cu-kα radiation on a Rigaku FR-X diffractometer equipped an Oxford Cryosystems nitrogen flow gas system. Details on the crystal structure determination and refinements can be found in Section S2 of the Supporting Information. CCDC 2068293-2068295 contains the supplementary crystallographic data for this paper. These data can be obtained free of charge via www.ccdc.cam.ac.uk/conts/retrieving.html (or from the Cambridge Crystallographic Data Centre, 12 Union Road, Cambridge CB21EZ, UK; fax: (+44)1223-336-033; or deposit@ccdc.cam.ac.uk).

**Magnetic properties.** Variable-temperature (2–300 K) direct current (d.c.) magnetic susceptibility measurements were carried out in an applied field of 1.0 kOe and variable field magnetization measurements up to ±5 T at 2.0 K. The susceptibility data were corrected from the diamagnetic contributions as deduced by using Pascal's constant tables. Variable-temperature (16–23 K) alternating current (ac) magnetic susceptibility measurements in a ±4.0 G oscillating field at frequencies in the range of 1–997 Hz were carried out in a zero dc field. All the measurements were performed with a SQUID magnetometer (Quantum Design MPMS-XL-5 & MPMS-XL-7).

**Laser interferometry.** The motion of the nanodrums was measured using a laser interferometry set-up, similar to the one reported in Ref.[27]. The sample is mounted on a xyz piezomotive nanopositioning stage inside the chamber of Montana Cryostation s50 dry optical cryostat. A blue diode laser ($\lambda$ = 405 nm), that is power-modulated by a Vector Network Analyser (VNA), is focused at the centre of the membrane and used to optothermally excite the membrane into motion at a given frequency. A red He-Ne laser ($\lambda$ = 632 nm) is used to read out the vibrations of the nanodrum, which are analyzed using a homodyne detection scheme and processed by the VNA. Laser spot diameter is in the order of 1 μm. It is checked that the resonance frequency changes due to laser heating are insignificant for all membranes.

**Peak-force AFM.** All peak-force AFM data is acquired at a constant 50 nN of applied peak force using a cantilever with a spring constant k = 22,8 Nm$^{-1}$ on Bruker Dimension FastScan AFM. The thickness of the sample is estimated taking the average of 3 to 5 profile scans.

## ASSOCIATED CONTENT

## AUTHOR INFORMATION


### Corresponding Author
*guillermo.minguez@uv.es, eugenio.coronado@uv.es

### Author Contributions
The manuscript was written through contributions of all authors. All authors have given approval to the final version of the manuscript.


## ACKNOWLEDGMENT


Financial support from the EU (ERC Advanced Grant MOL-2D 788222, ERC Consolidator Grant S-CAGE 724681 and FET-OPEN SINFONIA 964396), the Spanish MICINN (PID2020-117177GB-I00, PID2020-117152RB-I00 co-financed by FEDER ,and Unit of Excellence María de Maeztu CEX2019-000919-M), the Generalitat Valenciana (PROMETEO program and iDiFEDER/2020/063) is gratefully acknowledged. J.L.-C. acknowledges the Universitat de València for an "Atracció de Talent" fellowship. M.S., M.L., P.G.S. and H.S.J.v.d.Z. acknowledge funding from the European Union's Horizon 2020 research and innovation program under grant agreement number 881603..

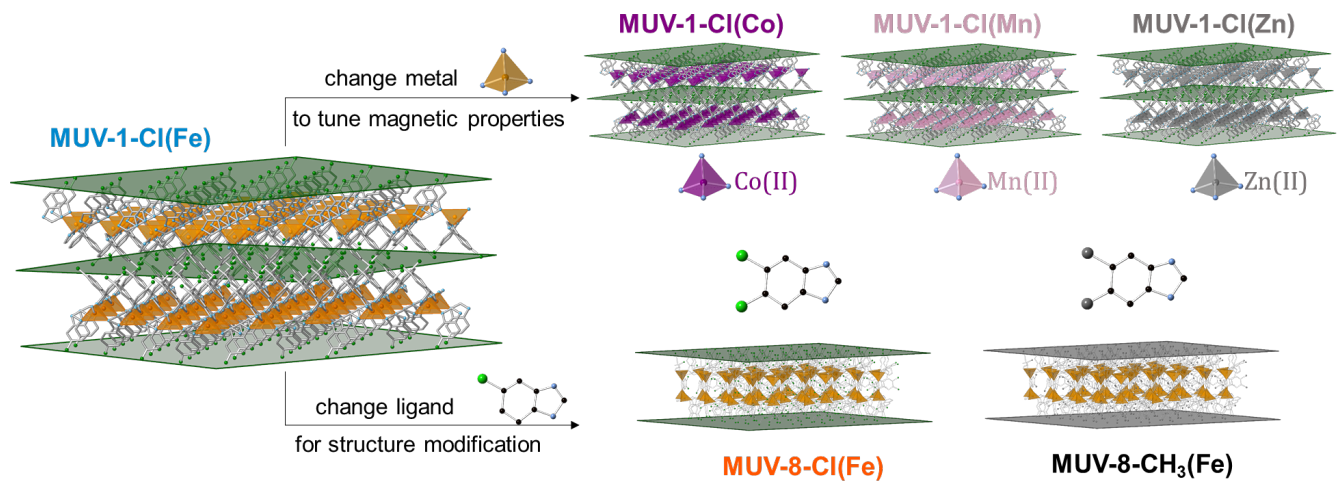



# Chemical design and magnetic ordering in thin layers of 2D MOFs


Javier López-Cabrelles,[$,1] Samuel Mañas-Valero,[$,1] Iñigo J. Vitórica-Yrezábal,[2] Makars Šiškins,[3] Martin Lee,[3] Peter G. Steeneken,[3,4] Herre S. J. van der Zant,[3] Guillermo Mínguez Espallargas*[1] and Eugenio Coronado*[1]

[1]Instituto de Ciencia Molecular (ICMol), Universidad de Valencia, c/Catedrático José Beltrán, 2, 46980 Paterna, Spain.

[2]School of Chemistry, University of Manchester, Manchester, UK.

[3]Kavli Institute of Nanoscience, Delft University of Technology, Lorentzweg 1, 2628 CJ, Delft, The Netherlands.

[4]Department of Precision and Microsystems Engineering, Delft University of Technology, Mekelweg 2, 2628 CD, Delft, The Netherlands.

[$]Equally contributed


# Contents





# S1. Materials and synthesis

**Synthesis.** The compounds were prepared adapting a previously described method for the preparation of iron imidazolates. All reagents and solvents were commercially available and used without further purification.

### Synthesis of MUV-1-Cl(Co) [Co(bim)$_2$]

Cobaltocene (30 mg, 0.16 mmol) and benzimidazole (40 mg, 0.34 mmol) were combined and sealed under vacuum in a layering tube (4 mm diameter). The mixture was heated at 150 °C for 4 days to obtain purple crystals suitable for X-ray single-crystal diffraction. The product was allowed to cool to room temperature, and the layering tube was then opened. The unreacted precursors were extracted with acetonitrile and benzene, and the main compound was isolated as purple crystals. Phase purity was established by X-ray powder diffraction.

### Synthesis of MUV-1-Cl(Co) [Co(bimCl)$_2$]

Cobaltocene (30 mg, 0.16 mmol) and 5-chlorobenzimidazole (49 mg, 0.34 mmol) were combined and sealed under vacuum in a layering tube (4 mm diameter). The mixture was heated at 150 °C for 4 days to obtain purple crystals suitable for X-ray single-crystal diffraction. The product was allowed to cool to room temperature, and the layering tube was then opened. The unreacted precursors were extracted with acetonitrile and benzene, and the main compound was isolated as purple crystals. Phase purity was established by X-ray powder diffraction.

### Synthesis of MUV-1-Cl(Mn) [Mn(bim)$_2$]

Bis(tetramethylcyclopentadienyl)manganese(II) (47.5 mg, 0.16 mmol) and benzimidazole (40 mg, 0.34 mmol) were combined and sealed under vacuum in a layering tube (4 mm diameter). The mixture



was heated at 250 °C for 3 days to obtain colourless crystals suitable for X-ray single-crystal diffraction. The product was cooled down naturally to room temperature, and the layering tube was then opened. The unreacted precursors were extracted with acetonitrile and benzene, and the main compound was isolated as colourless crystals. Phase purity was established by X-ray powder diffraction.

**Synthesis of MUV-1-Cl(Mn) [Mn(bimCl)$_2$]**

Bis(tetramethylcyclopentadienyl)manganese(II) (47.5 mg, 0.16 mmol) and 5-chlorobenzimidazole (49 mg, 0.34 mmol) were combined and sealed under vacuum in a layering tube (4 mm diameter). The mixture was heated at 250 °C for 3 days to obtain colourless crystals suitable for X-ray single-crystal diffraction. The product was allowed to cool to room temperature, and the layering tube was then opened. The unreacted precursors were extracted with acetonitrile and benzene, and the main compound was isolated as colourless crystals. Phase purity was established by X-ray powder diffraction.

**Synthesis of MUV-1-H(Zn) [Zn(bim)$_2$]**

Bis(2,2,6,6-tetramethyl-3,5-heptanedionato) zinc(II) (69.1 mg, 0.16 mmol) and benzimidazole (40 mg, 0.34 mmol) were combined and sealed under vacuum in a layering tube (4 mm diameter). The mixture was heated at 180 °C for 4 days to obtain colourless crystals suitable for X-ray single-crystal diffraction. The product was allowed to cool to room temperature, and the layering tube was then opened. The unreacted precursors were extracted with acetonitrile and benzene, and the main compound was isolated as colourless crystals. Phase purity was established by X-ray powder diffraction.



**Synthesis of MUV-1-Cl(Zn) [Zn(bimCl)$_2$]**

Bis(2,2,6,6-tetramethyl-3,5-heptanedionato) zinc(II) (69.1 mg, 0.16 mmol) and 5-chlorobenzimidazole (49 mg, 0.34 mmol) were combined and sealed under vacuum in a layering tube (4 mm diameter). The mixture was heated at 180 °C for 4 days to obtain colourless crystals suitable for X-ray single-crystal diffraction. The product was allowed to cool to room temperature, and the layering tube was then opened. The unreacted precursors were extracted with acetonitrile and benzene, and the main compound was isolated as colourless crystals. Phase purity was established by X-ray powder diffraction.

**Synthesis of MUV-8-Cl(Fe) [Fe(bimCl$_2$)$_2$]**

Ferrocene (30 mg, 0.16 mmol) and 5,6-dichlorobenzimidazole (64 mg, 0.34 mmol) were combined and sealed under vacuum in a layering tube (4 mm diameter). The mixture was heated at 250 °C for 3 days to obtain colourless crystals suitable for X-ray single-crystal diffraction. The product was allowed to cool to room temperature, and the layering tube was then opened. The unreacted precursors were extracted with acetonitrile and benzene, and the main compound was isolated as colourless crystals. Phase purity was established by X-ray powder diffraction.

**Synthesis of MUV-8-CH$_3$(Fe) [Fe(bim(CH$_3$)$_2$)$_2$]**

Ferrocene (30 mg, 0.16 mmol) and 5,6-dimethylbenzimidazole (50 mg, 0.34 mmol) were combined and sealed under vacuum in a layering tube (4 mm diameter). The mixture was heated at 250 °C for 3 days to obtain colourless crystals suitable for X-ray single-crystal diffraction. The product was allowed to cool to room temperature, and the layering tube was then opened. The unreacted precursors were extracted with acetonitrile and benzene, and the main compound was isolated as colourless crystals (yield 60 %). Phase purity was established by X-ray powder diffraction.



# S2. Structure of MUV-1-X(M(II)) and MUV-8-X(Fe) family

## S2.1 Data Collection

X-Ray data for compound **MUV-1-H(Co) and MUV-8-Cl(Fe)** were collected at a temperature of 120 K using a Mo-kα radiation on a Rigaku Supernova diffractometer equipped an Oxford Cryosystems nitrogen flow gas system. X-Ray data for compound **MUV-1-Cl(Mn)** were collected at a temperature of 100 K using a Cu-kα radiation on a Rigaku FR-X diffractometer equipped an Oxford Cryosystems nitrogen flow gas system. Data were measured using CrisAlisPro suite of programs.

## S2.2 Crystal structure determinations and refinements

X-Ray data were processed and reduced using CrysAlisPro suite of programmes. Absorption correction was performed using empirical methods (SCALE3 ABSPACK) based upon symmetry-equivalent reflections combined with measurements at different azimuthal angles. The crystal structure was solved and refined against all $F^2$ values using the SHELXL and Olex 2 suite of programmes.[1,2] All atoms were refined anisotropically. Hydrogen atoms were placed in the calculated positions. Pyridine moieties in crystal **MUV-1-H(Co)** were disordered and modelled over two positions. Ferrocene molecules were disordered and modelled over two positions in **MUV-8-Cl(Fe)**. The atomic displacement parameters were also restrained using SHELXs SIMU and RIGU commands.

Crystals of **MUV-1-H(Co)** and **MUV-1-Cl(Mn)** were found to be modulated commensurate with a q vector (½, ½, 0). Despite that the coordination polymers are intrinsically chiral; the centrosymmetric space group C2/c was found as result of the racemic distribution of the disordered layers.



CCDC 2068293-2068295 contains the supplementary crystallographic data for this paper. These data can be obtained free of charge *via* www.ccdc.cam.ac.uk/conts/retrieving.html (or from the Cambridge Crystallographic Data Centre, 12 Union Road, Cambridge CB21EZ, UK; fax: (+44)1223-336-033; or deposit@ccdc.cam.ac.uk).

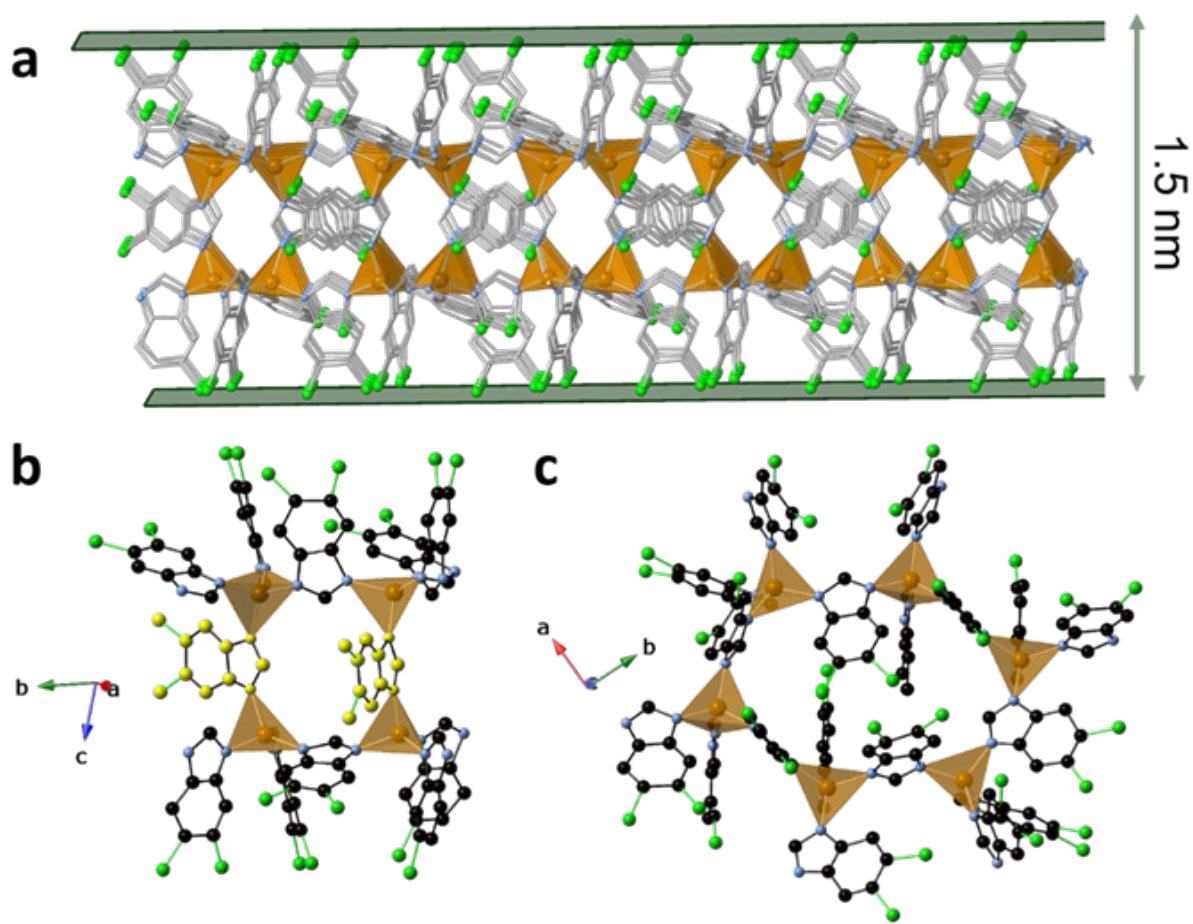

**Figure S1**. a) Crystal structure of the **MUV-8-Cl(Fe)**, composed by Fe(II) centers in a tetrahedral environment, forming extended layers in the *ab* plane. b) The connectivity in one sheet, three ligands connecting in *ab* plane, and the fourth ligand (in yellow) connecting the two layers of iron centers (the Fe bilayer). c) The distorted hexagonal lattice in the perpendicular *ab* plane view.



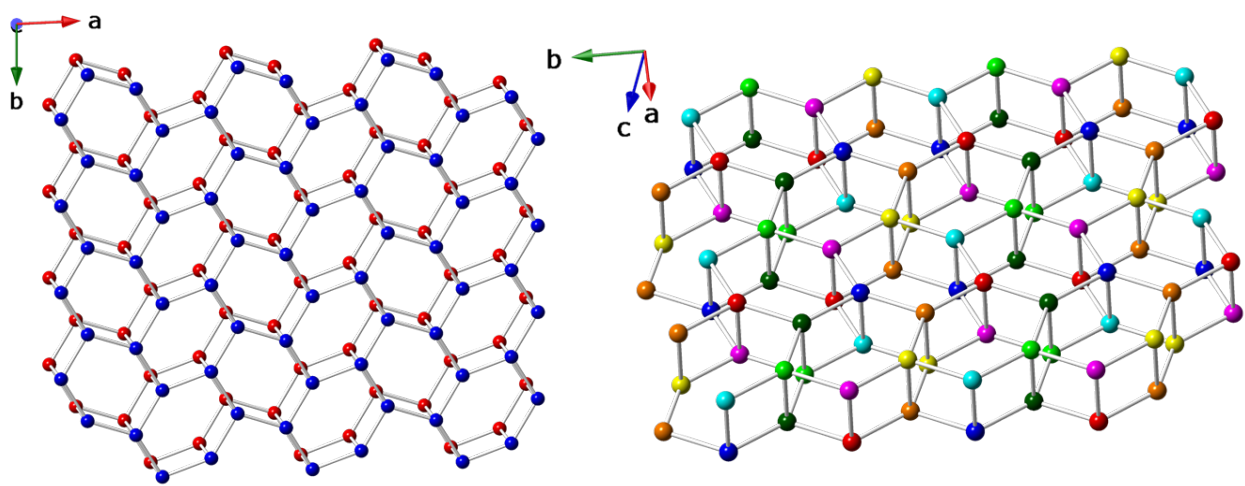

**Figure S2.** Crystal structure of **MUV-8-Cl(Fe).** Bilayered structure for a monolayer, similar to AB stacking. Only iron centres are represented (C, N, H, and Cl atoms are omitted) (left). Seven different iron centres in a monolayer. Different colours have been used to differentiate one from the other. Only iron centres are represented (C, N, H, and Cl atoms are omitted) (right).

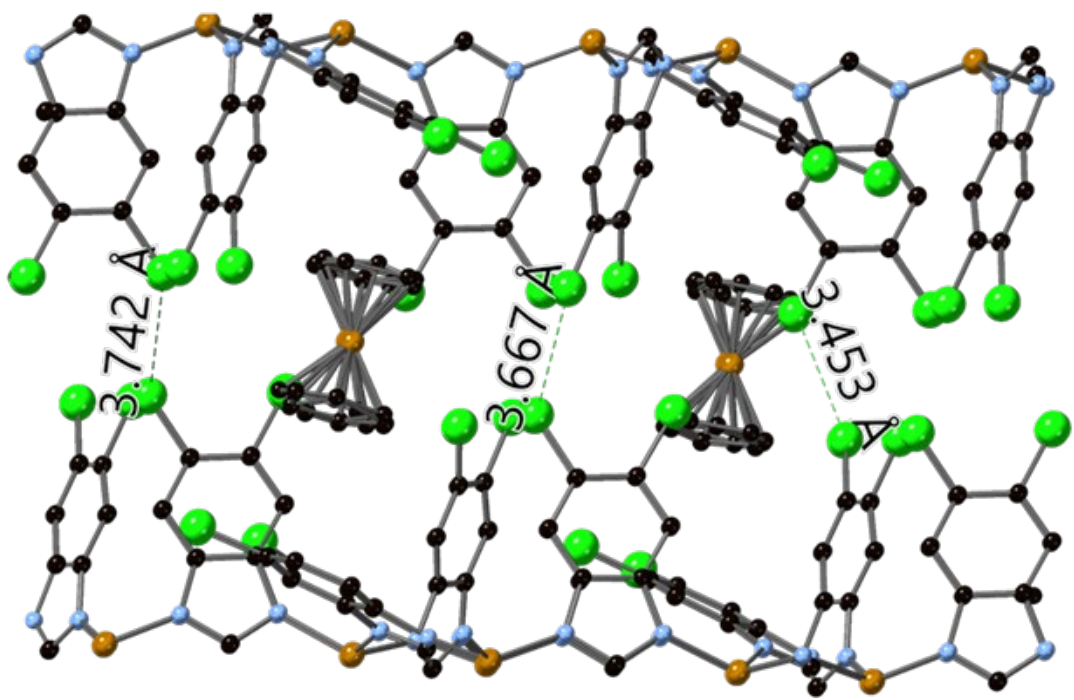

**Figure S3.** Crystal structure of **MUV-8-Cl(Fe)**. with the location of ferrocene molecules in the interlayer space.



**Table S1.** Crystallographic information for compounds **MUV-1-X** and **MUV-8-Cl(Fe).**

| Identification code | **MUV-8-Cl(Fe)** | **MUV-1-H(Co)** | **MUV-1-Cl(Mn)** |
|---|---|---|---|
| Empirical formula | $C_{66}H_{34}Cl_{16}Fe_5N_{16}$ | $C_{14}H_{10}CoN_4$ | $C_{14}H_8Cl_2MnN_4$ |
| Formula weight | 1897.54 | 293.19 | 358.08 |
| Temperature/K | 120.00(10) | 119.8(4) | 99.8(8) |
| Crystal system | triclinic | monoclinic | monoclinic |
| Space group | P-1 | C2/c | C2/c |
| a/Å | 10.4429(6) | 8.0785(11) | 8.3680(2) |
| b/Å | 17.5917(9) | 8.0783(12) | 8.3838(2) |
| c/Å | 21.340(2) | 19.258(3) | 20.9429(6) |
| α/° | 100.270(7) | 90 | 90 |
| β/° | 101.992(7) | 96.159(13) | 95.897(2) |
| γ/° | 90.045(4) | 90 | 90 |
| Volume/Å³ | 3770.4(5) | 1249.5(3) | 1461.49(6) |
| Z | 2 | 4 | 4 |
| $\rho_{calc}$ g/cm³ | 1.671 | 1.559 | 1.627 |
| μ/mm⁻¹ | 1.559 | 1.362 | 10.684 |
| F(000) | 1888.0 | 596.0 | 716.0 |
| Crystal size/mm³ | 0.1 × 0.06 × 0.05 | 0.1 × 0.1 × 0.1 | 0.03 × 0.02 × 0.02 |
| Radiation | Mo Kα (λ = 0.71073) | Mo Kα (λ = 0.71073) | Cu Kα (λ = 1.54184) |
| 2Θ range for data collection/° | 3.344 to 50.7 | 7.154 to 50.038 | 8.49 to 151.912 |
| Index ranges | -12 ≤ h ≤ 12, -21 ≤ k ≤ 21, -25 ≤ l ≤ 25 | -9 ≤ h ≤ 9, -8 ≤ k ≤ 9, -3 ≤ l ≤ 22 | -10 ≤ h ≤ 10, -10 ≤ k ≤ 10, -25 ≤ l ≤ 23 |
| Reflections collected | 13813 | 1108 | 8674 |
| Independent reflections | 13813 [$R_{int}$ = 0.1984, $R_{sigma}$ = 0.2133] | 1108 [$R_{int}$ = 0.101, $R_{sigma}$ = 0.0766] | 1525 [$R_{int}$ = 0.0450, $R_{sigma}$ = 0.0334] |
| Data/restraints/parameters | 13813/1393/1029 | 1108/222/152 | 1525/0/168 |
| Goodness-of-fit on F² | 1.041 | 1.263 | 1.036 |
| Final R indexes [I>=2σ (I)] | $R_1$ = 0.0967, $wR_2$ = 0.2102 | $R_1$ = 0.1193, $wR_2$ = 0.2820 | $R_1$ = 0.0400, $wR_2$ = 0.1008 |
| Final R indexes [all data] | $R_1$ = 0.1991, $wR_2$ = 0.2647 | $R_1$ = 0.1262, $wR_2$ = 0.2861 | $R_1$ = 0.0413, $wR_2$ = 0.1025 |
| Largest diff. peak/hole / e Å⁻³ | 1.70/-1.14 | 1.55/-3.29 | 0.35/-0.48 |

[a] $R1(F) = \Sigma(|F_o| - |F_c|)/\Sigma|F_o|$; [b] $wR^2(F^2) = [\Sigma w(F_o^2 - F_c^2)^2/\Sigma wF_o^4]^{½}$; [c] $S(F^2) = [\Sigma w(F_o^2 - F_c^2)^2/(n + r - p)]^{½}$



# S3. Characterization of MUV-1-X(M(II) = Co, Mn, Zn) family and MUV-8-X(Fe)

Polycrystalline samples of **MUV-1-H(Co), MUV-1-Cl(Co), MUV-1-H(Mn), MUV-1-Cl(Mn), MUV-8-Cl(Fe)**, **MUV-8-CH$_3$(Fe)** were lightly ground in an agate mortar and pestle and used to fill a 0.5 mm borosilicate capillariy that was mounted and aligned on an Empyrean PANalytical powder diffractometer, using Cu Kα radiation (λ = 1.54056 Å). Two repeated measurements were collected at room temperature (2θ = 5−40 °) and merged in a single diffractogram. Thermogravimetric analysis of **MUV-1-H(Co)** and **MUV-1-Cl(Co)** were carried out with a Mettler Toledo TGA/SDTA851e/SF/1100 apparatus in the 25–600 °C temperature range under a 5°C·min$^{-1}$ scan rate and an air flow of 30 mL·min$^{-1}$. Scanning Electronic Micrographs and atomic composition of bulk samples was estimated by electron probe microanalysis (EPMA) performed in a Philips SEM XL30 equipped with an EDAX microprobe and images were recorded in a Hitachi S-4800. Raman spectra were acquired with a micro-Raman (model XploRA ONE from Horiba, Kyoto, Japan) with a grating of 2400 gr/mm, slit of 50 μm, and hole of 500 μm. The employed wavelengths were 532 nm, 638 nm, and 785 nm. The power density of the laser used for spectra measured at 532 nm was 5.25 mW/μm$^2$ (bulk crystals) and 170 μW/μm$^2$ (thin-layers), for spectra measured at 638 nm it was 7.58 mW/μm$^2$ (bulk crystals), and for those spectra measured at 785 nm it was 7.2 mW/μm$^2$ (bulk crystals).

**S3.1 X-Ray powder diffraction**

In Figure S4-S5 the experimental X-ray powder diffraction patterns (in color) are summarized and compared with the theoretical diffractogram. It is worth to mention the high correlation between the experimental and the theoretical diffractograms.



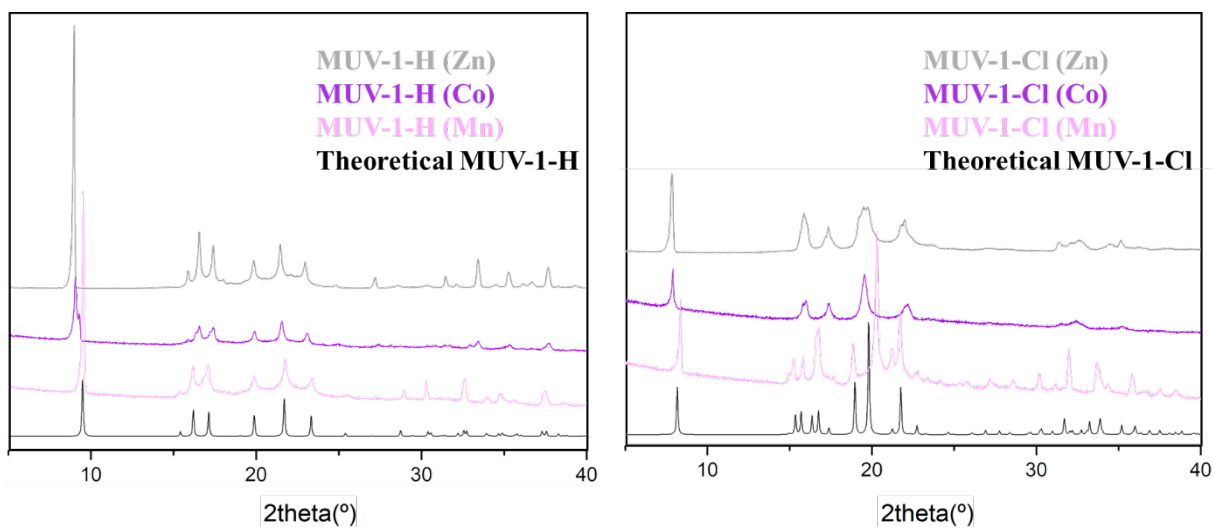

**Figure S4.** X-ray powder patterns of **MUV-1-H(M(II))** and **MUV-1-Cl(M(II)).** The experimental patterns are shown in pink (Mn(II)), purple (Co(II)) and grey (Zn(II)) for the respective compounds, and the calculated pattern from single crystal data are shown in black for all compounds.

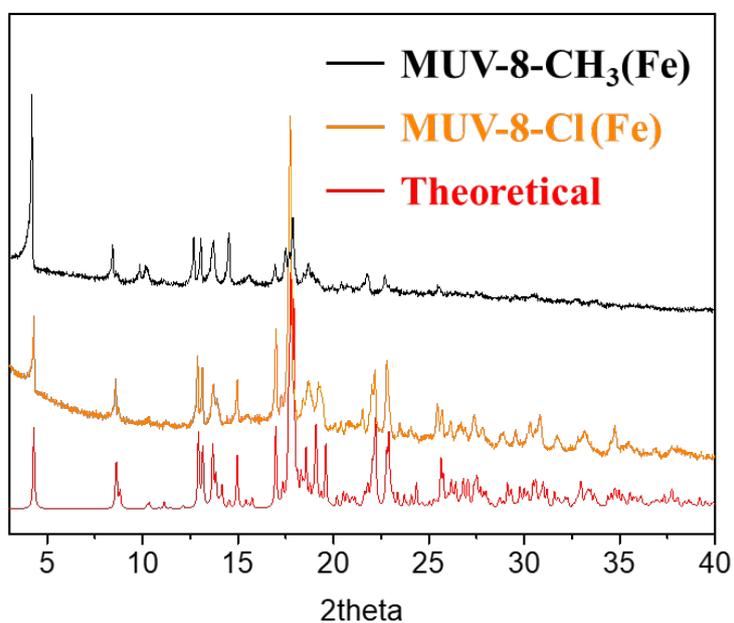

**Figure S5.** X-ray powder patterns of **MUV-8-Cl(Fe)** and **MUV-8-CH$_3$(Fe)**. The experimental patterns are shown in orange and black for the respective compounds, and the calculated pattern from single crystal data are shown in red for all compounds.



## S3.2 Thermal stability

The thermal stability of all the compounds was determined by thermal gravimetric analysis at a heating rate of 20 °C min$^{-1}$.

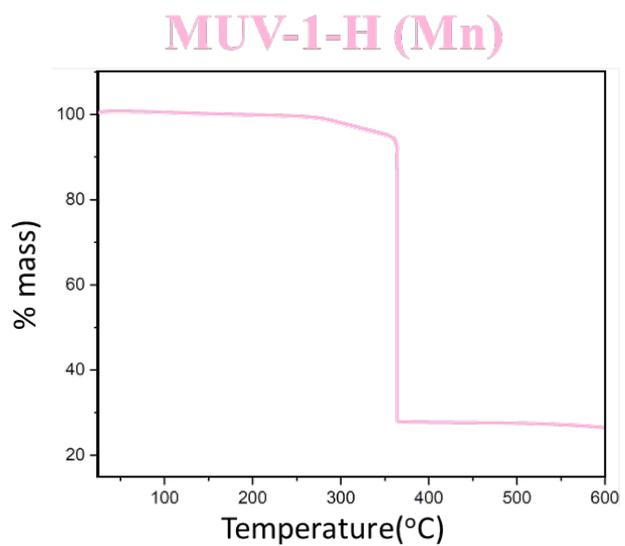
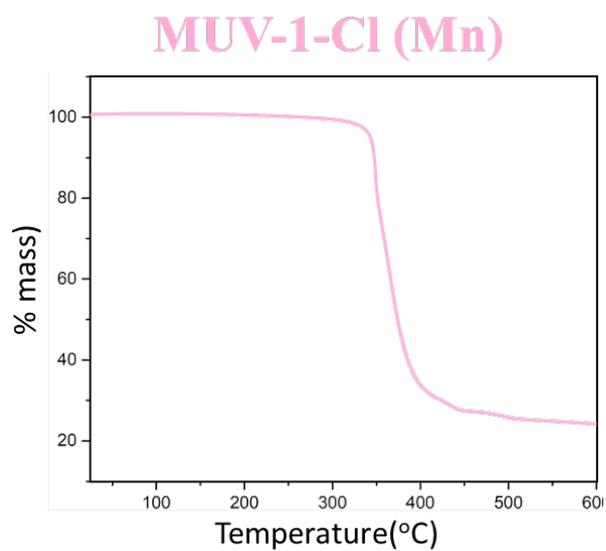
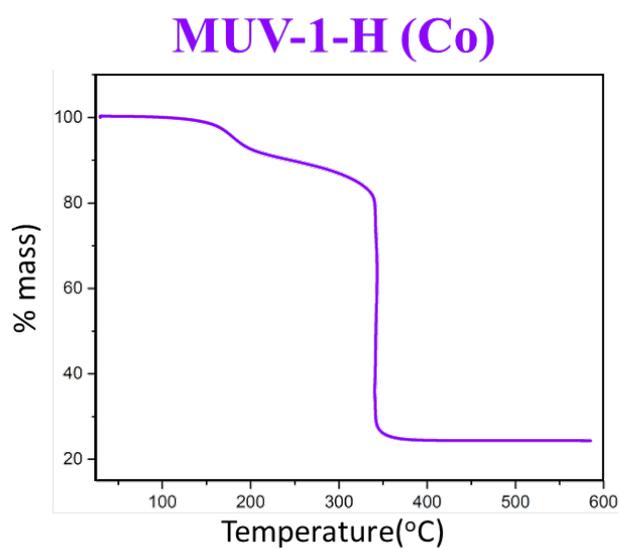
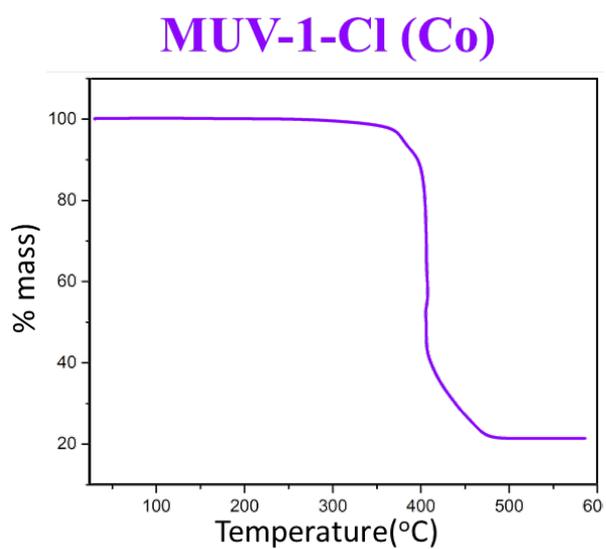



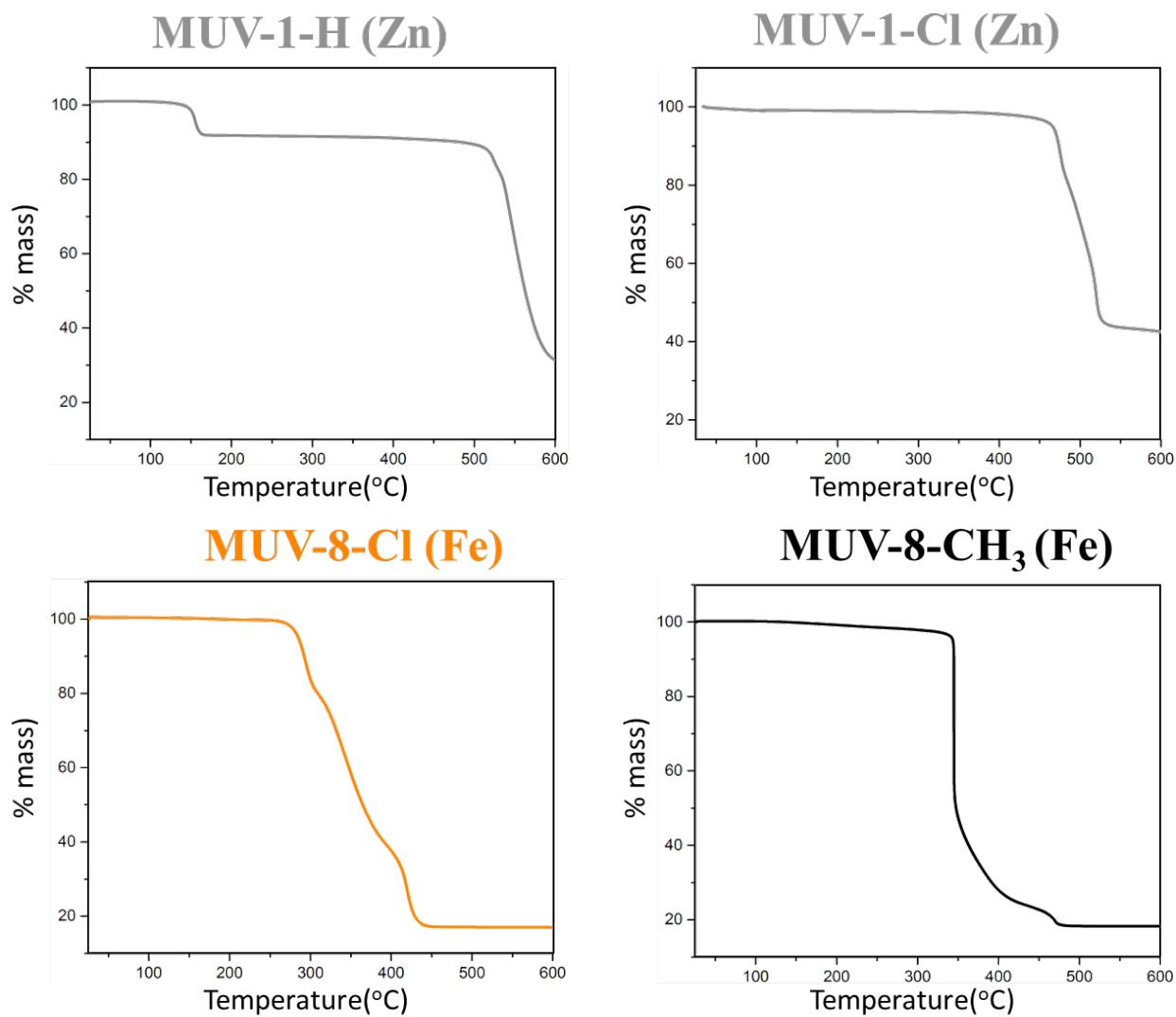

**Figure S6.** Thermal gravimetric analysis (TGA) of bulk crystals of **MUV-1-H(Co), MUV-1-Cl(Co)**, **MUV-1-H(Mn), MUV-1-Cl(Mn), MUV-1-H(Zn), MUV-1-Cl(Zn), MUV-8-Cl(Fe)** and **MUV-8-CH$_3$(Fe)** at a heating rate of 20 ºC min$^{-1}$.



## S3.3 Scanning electronic microscope images

The layered structure of **MUV-1-Cl(Co), MUV-1-H(Co), MUV-1-Cl(Mn) and MUV-1-H(Mn)**, was clearly identified by SEM images. As it can be seen in Figure S7-S10, well-defined rectangular edges are the main characteristic of these compounds.

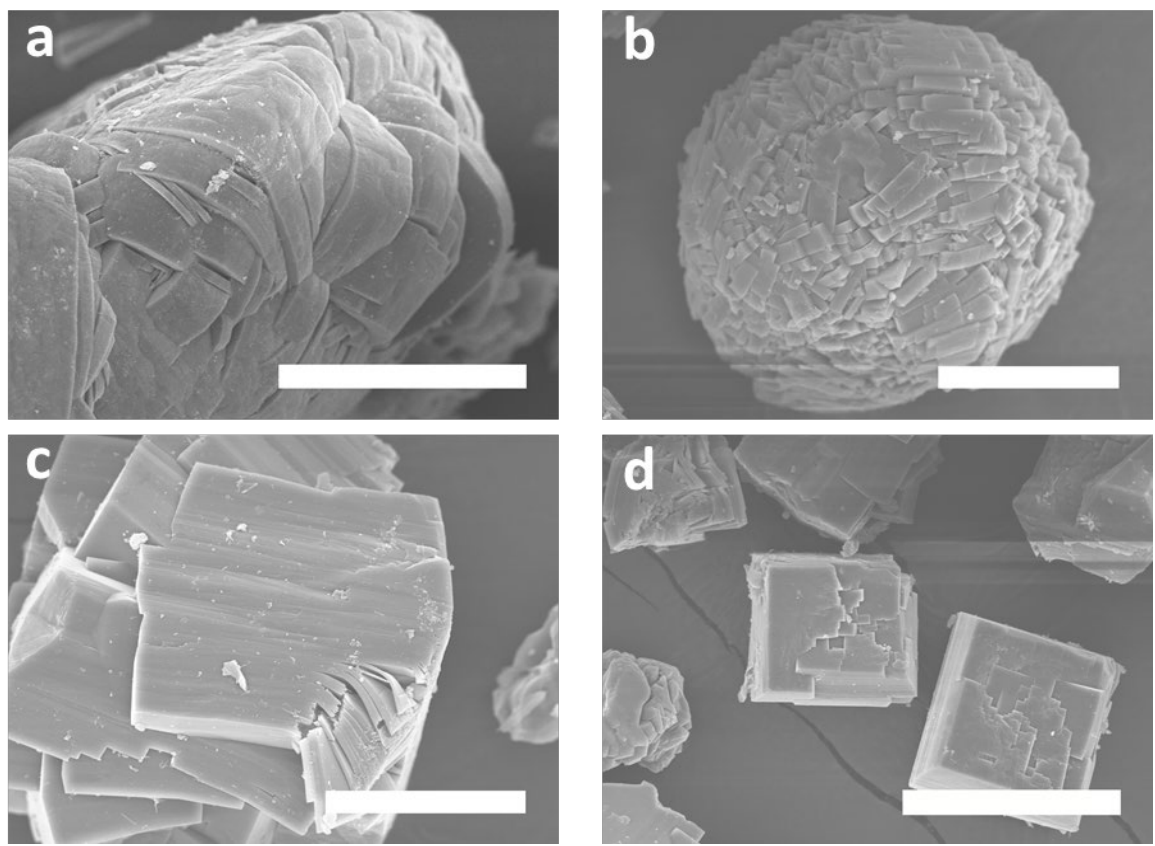

**Figure S7.** Scanning electron micrograph of bulk-type **MUV-1-Cl(Co)** (a,b) and **MUV-1-H(Co)** (c,d). Scale bar is a) 50 µm b) 20 µm c) 20 µm d) 40 µm.



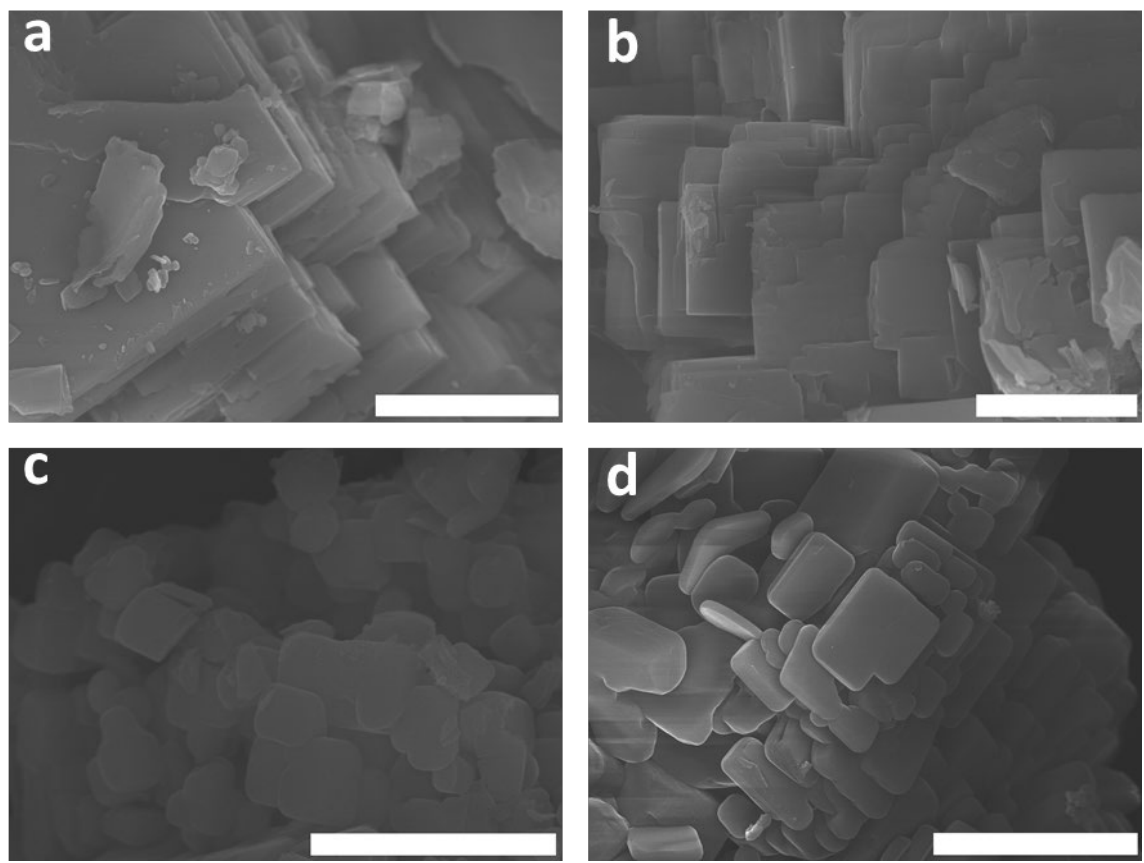

**Figure S8**. Scanning electron micrograph of bulk-type **MUV-1-Cl(Mn)** (a,b) and **MUV-1-H(Mn)** (c,d)**.** Scale bar is a-b) 10 µm c-d) 5 µm.



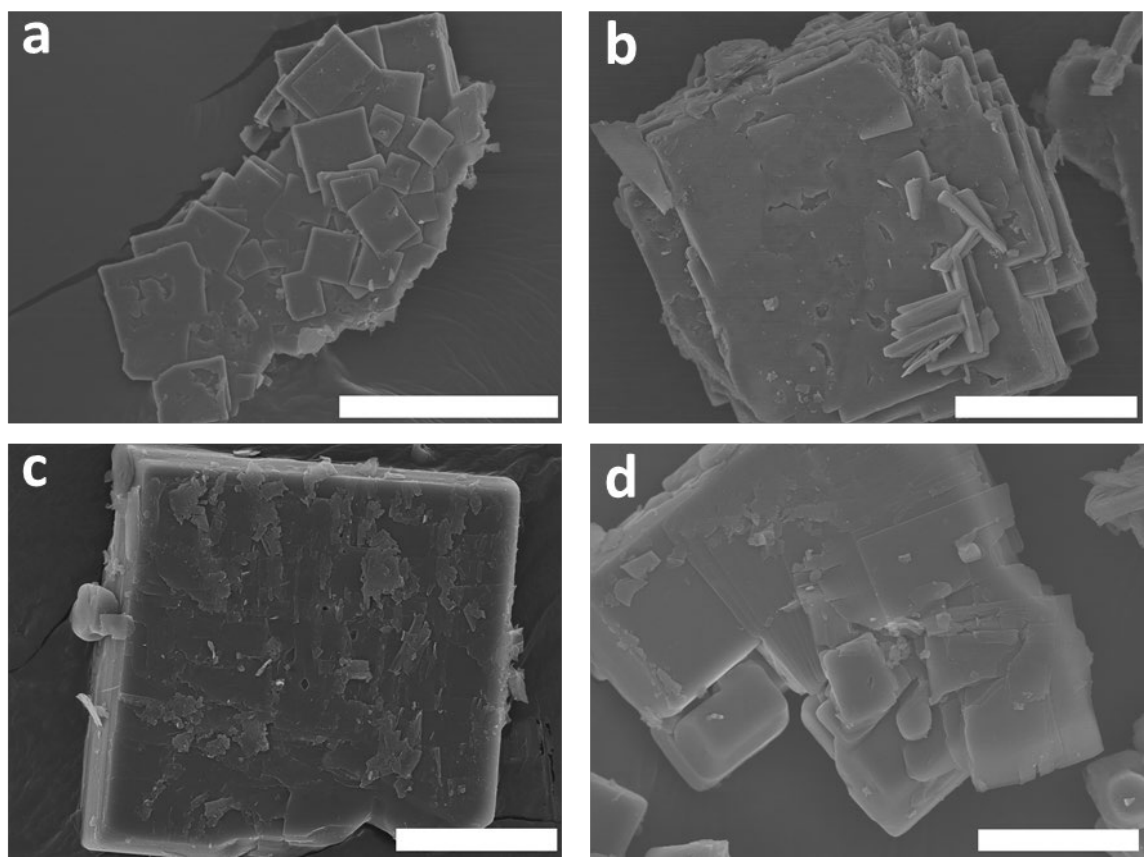

**Figure S9**. Scanning electron micrograph of bulk-type **MUV-1-Cl(Zn)** (a,b) and **MUV-1-H(Zn)** (c,d)**.** Scale bar is a) 40 μm b) 20 μm c-d) 10 μm.



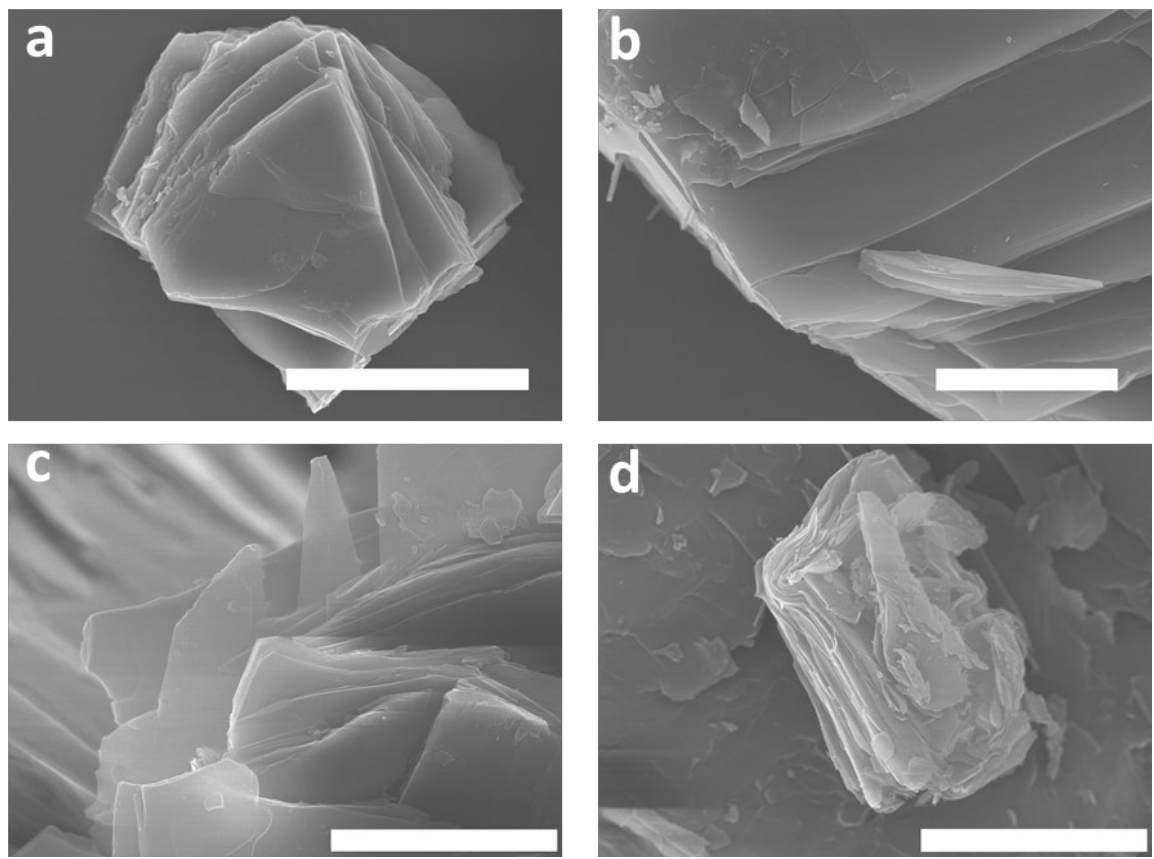

**Figure S10.** Scanning electron micrograph of bulk-type **MUV-8-Cl(Fe)** (a,b) and **MUV-8-CH3(Fe)** (c,d). Scale bar is a) 30 μm b-d) 10.



# S4. Magnetic properties

## S4.1 SQUID measurements and magnetic models

Magnetic measurements are performed with a SQUID magnetometer (Quantum Design MPMS-XL-5). Direct current (dc) magnetic susceptibility is measured with an applied field of 1000 Oe whereas alternating current (ac) measurements are performed at different frequencies in a zero dc field and an oscillating 4 G field. Electron Paramagnetic Resonance (EPR) spectroscopy measurements were recorded with a Bruker ELEXYS E580 spectrometer operating in the Q band (34 GHz).

The magnetic measurements for all the different compounds are shown in Figure S13-S14 (magnetic susceptibility, $\chi$, versus temperature), Figure S16 ($\chi T$ vs. T), Figure S.17 ($\chi^{-1}$ vs. T), Figure S18 (magnetization vs. external field at 2 K) and Figure S19-S21 (a.c. $\chi$ vs. T).

The temperature dependence of the dc susceptibility increases while cooling down until a broad maximum is reached. Below this temperature, there is a sharp increase of $\chi$ –attributed to spin-canting and that it is accompanied by a sharp peak in the ac measurements– for all the compounds except for the manganese systems followed by a decrease of the susceptibility due to the existent antiferromagnetic interactions. In some compounds, it is observed an enhancement of the susceptibility at the lowest temperatures due to the presence of some paramagnetic impurities. The Néel temperature is assigned as the point where the out-of-phase component of the a.c. susceptibility differs from 0 in the case of the systems with spin-canting and following the Fisher criteria (maximum in $\frac{\partial(\chi T)}{\partial T}$) in the case where the canting is absent (**MUV-1(Mn)**).[3,4]

In the case of the **MUV-1** family, the magnetic coupling constant, J, is determined by fitting the susceptibility versus temperature curves with the Lines expression[16] for a quadratic-layer antiferromagnet based on the following Hamiltonian: $H = -\sum_{i,j} J S_i \cdot S_j$, where J < 0 indicates antiferromagnetic correlations.



$$\chi = -\frac{Ng^2\mu_B^2}{J}\left(3\frac{k_BT}{JS(S+1)} + \sum_{n=1}^{6}\frac{C_n}{\left(\frac{k_BT}{JS(S+1)}\right)^{n-1}}\right)^{-1}$$

where J is the exchange constant, $k_B$ is the Boltzmann constant, S is the spin, N is the Avogadro´s number, g is the Landé factor, $\mu_B$ is the Bohr magneton and $C_n$ are coefficients of the Lines expansion (see ref. [16]).

Regarding the **MUV-8-X(Fe)** compounds, the magnetic properties are more challenging to model due to their complex topology. As an initial ansatz, we approximate it to a 2D Heisenberg honeycomb lattice with antiferromagnetic correlations between classical spins, as developed by Curély et al.[17] This model adapts the two-dimensional Heisenberg classical square lattice case to the hexagonal one by considering two different exchange constants, J and $J_0$. The former is related to vertical couplings while the latter takes into account the horizontal ones,[17] as sketched in Figure S11. In the limiting cases, the model yields to the formation of chains ($J_0 = 0$) or dimers (J = 0). The model starts with a square lattice with four different exchange constants between first neighbor atoms (left panel in Figure S11). By setting one of them to zero and equaling the perpendicular ones (J'$_0$ = 0 and J = $J_1$ in the middle panel of Figure S11), a hexagonal arrangement is obtained, which is more evident if one exchange coupling is stronger than the other, as sketched in the right panel of Figure S11. We note that, as well, there could be a possible exchange pathway between the iron centers forming the double layer (denoted as $J_{inter}$ in Figure S12). Evidently, the Curély model does not take into account this interaction. In a first approximation, we consider $J_{inter} = 0$ since the distance between the iron centers in the out-of-plane direction (6.178 Å – 6.184 Å) is larger than in the in-plane one (5.986 Å – 6.089 Å).

Following the Curély expression, the susceptibility for a compensated lattice takes the form:

$$\chi = \frac{\beta}{6}\frac{(G^2 + G'^2)W_1 + 2GG'W_2}{(1 - u^2v^2)(1 - v^2)}$$



where G is the associated Landé factor (in $\mu_B/\hbar$ units), $\beta=1/k_BT$ and with

$$W_1 = (1 + uv)^2(1 + v^2)$$

$$W_2 = 2v(1 + uv)^2 + u(1 - v^2)^2$$

u and v are given by the well-known Langevin function $\left(\mathcal{L}(x) = \coth(x) - \frac{1}{x}\right)$:

$$u = \mathcal{L}(-\beta J_0)$$

$$v = \mathcal{L}(-\beta J)$$

In the present case, we consider $J = J_0$ since J and $J_0$ are correlated (see Figure S11), thus not being possible to determine unambiguously J and $J_0$.

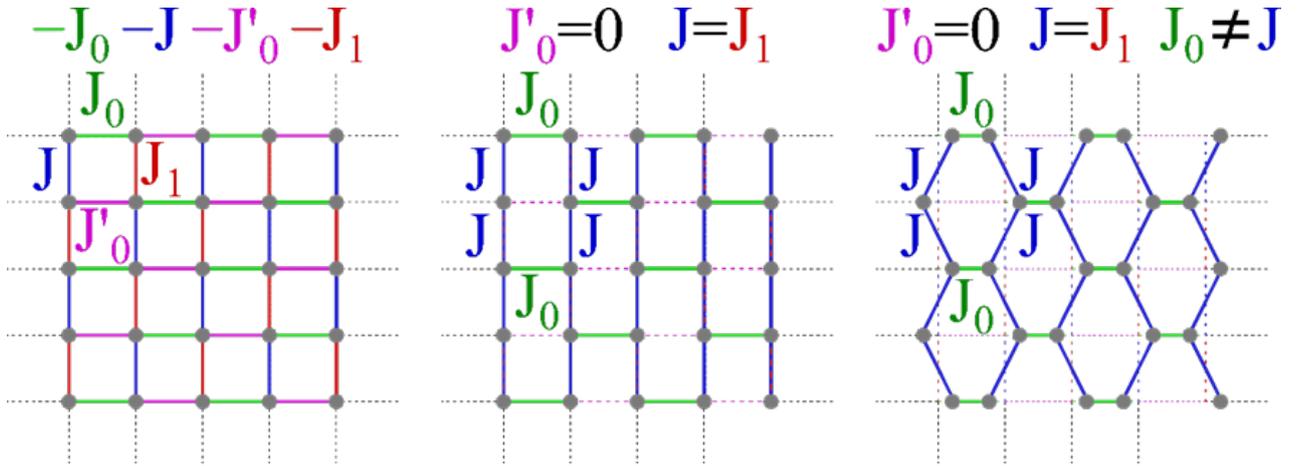

**Figure S.11**. Example of a transformation from a 2D lattice composed of classical spins and characterized by a square unit cell (left panel) to a hexagonal honeycomb lattice (middle and right panels).



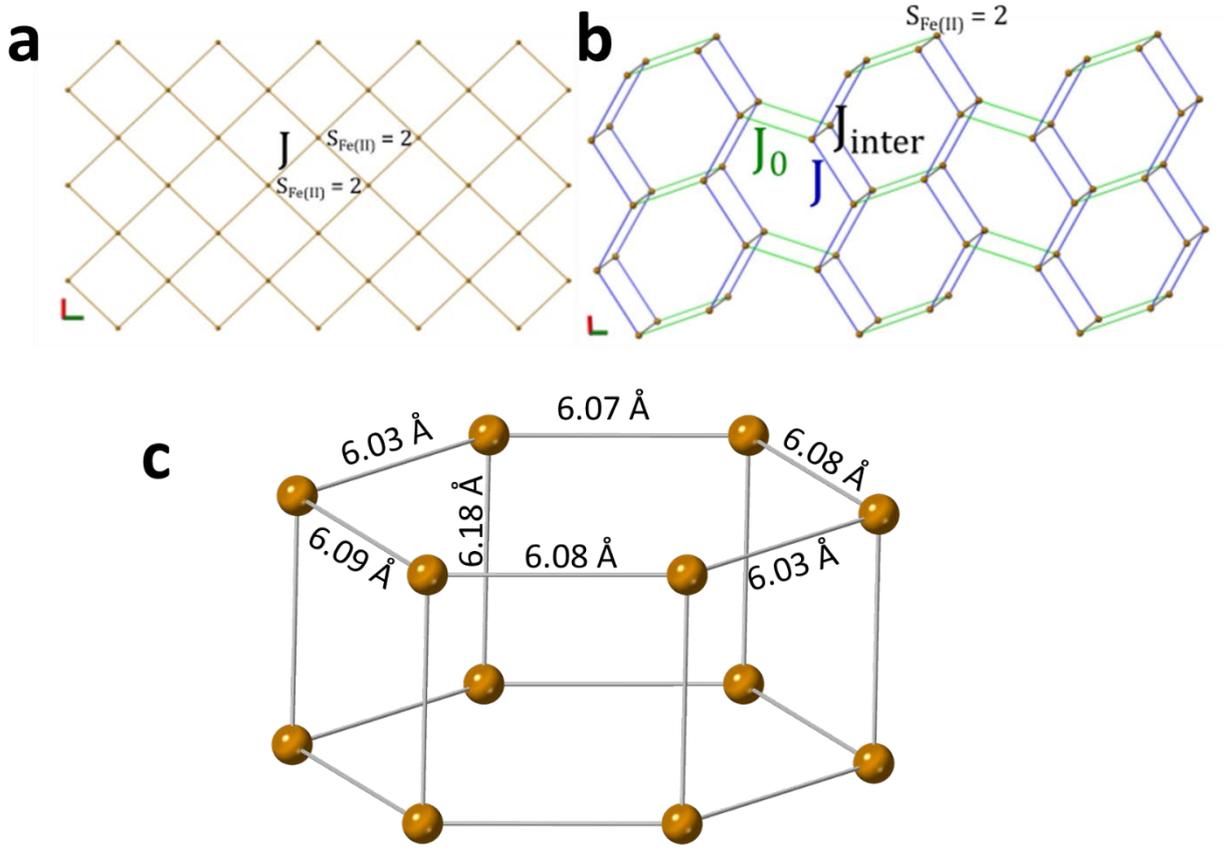

**Figure S.12**. Magnetic network for **MUV-1** (a) and **MUV-8** (b) compounds. Red, green and blue axis represent the a, b and c axis, respectively. c)Fe-Fe distances in the distorted hexagonal lattice.

Following the Curély formula (considering $H = \sum_{i,j} J S_i \cdot S_j$),[6] the susceptibility for a compensated lattice takes the form:

$$\chi = -\frac{\beta}{6} \frac{(G^2 + G'^2)W_1 + 2GG'W_2}{(1 - u^2v^2)(1 - v^2)}$$

where G is associated Landé factor (in $\mu_B/\hbar$ units), $\beta = 1/kT$ and with

$$W_1 = (1 + uv)^2(1 + v^2)$$

$$W_2 = 2v(1 + uv)^2 + u(1 - v^2)^2$$



u and v are given by the well-known Langevin function:

$$u = \mathcal{L}(-\beta J_0)$$

$$v = \mathcal{L}(-\beta J)$$

The obtained values for J and g are summarized in Table S.2, considering $J = J_0$. We note that, in the present case, J and $J_0$ are correlated, as can be seen in Figure S11.

In all compounds, the exchange coupling is antiferromagnetic as denoted by J < 0. In the case of the spin-canted systems, the J values are in consonance with the previous reported values for **MUV-1 (Fe)**. In the case of **MUV-1(Co)**, the exchange constant values are lower and, in accordance, they exhibit a lower Néel temperature. **Regarding MUV-1(Mn)**, where no spin-canting is observed, the J values are lower. Finally, **MUV-8(Fe)** exhibits similar Néel temperatures as **MUV-1(Fe)** thus, with similar values of J.

**Table S.2.-** Summary of the magnetic properties for the different compounds. J and g values for **MUV-1** and **MUV-8** are obtained according to the Lines and Curély expressions, respectively.[5,6]

| Structure type | $T_N$ (K) | J (cm$^{-1}$) | g |
|---|---|---|---|
| MUV-1-Cl(Fe) From ref. [7] | 20.7 | - 22.9 ± 0.4 | 2.000 ± 0.015 |
| MUV-1-H(Fe) From ref. [7] | 20.0 | - 23.5 ± 0.2 | 1.977 ± 0.011 |
| MUV-1-Cl(Co) | 11.6 | - 20.2 ± 0.4 | 2.3 ± 0.2 |
| MUV-1-H(Co) | 12.4 | - 20.8 ± 0.4 | 2.19 ± 0.15 |
| MUV-1-Cl(Mn) | 14.3 | - 10.7 ± 0.2 | 2.2 ± 0.2 |
| MUV-1-H(Mn) | 14.8 | - 10.41 ± 0.06 | 2.02 ± 0.09 |
| MUV-8-Cl(Fe) | 23.2 | - 19.1 ± 0.3 | 1.97 ± 0.18 |
| MUV-8-CH$_3$(Fe) | 23.4 | -25.2 ± 0.5 | 2.1 ± 0.2 |

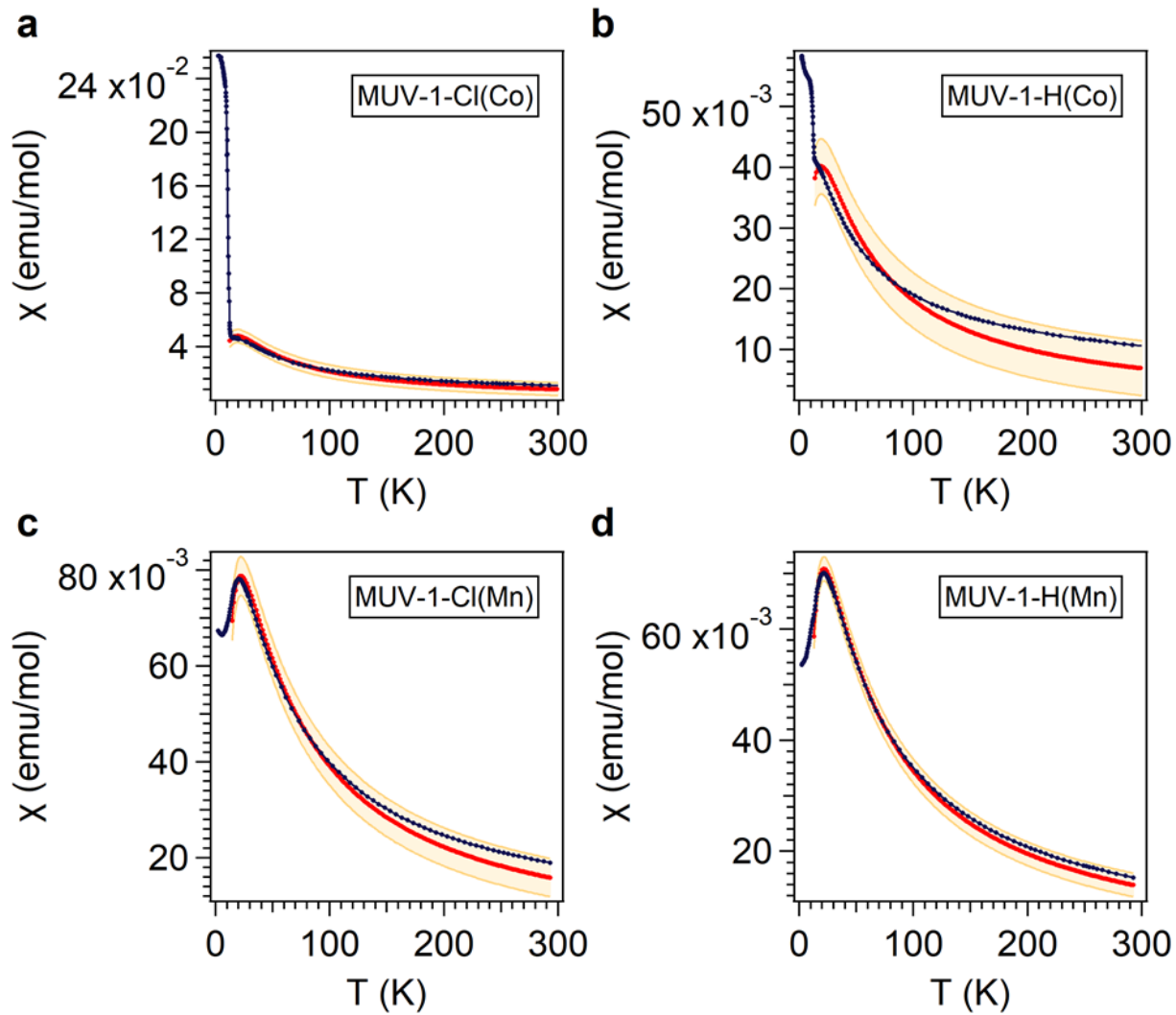

**Figure S13**. Determination of TN for **MUV-1(Mn)** following the Fisher criteria.[4]

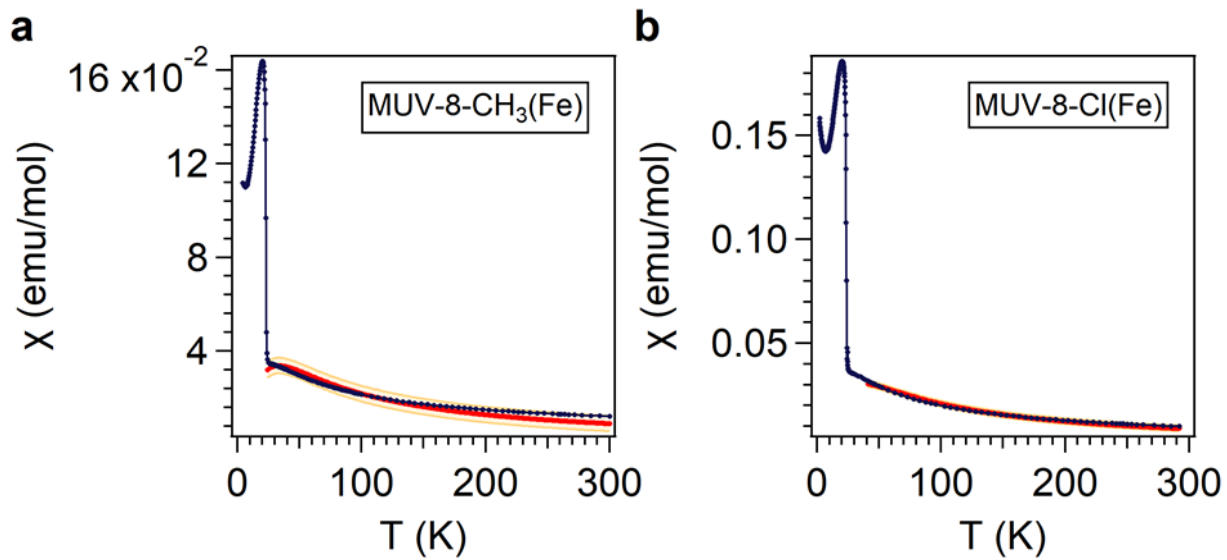

**Figure S14**. Thermal dependence of $\chi_m$ in the temperature range 2-300 K. The data have been fitted following the Curély analysis (solid red line).[6] It is represented in yellowish the prediction bands with a confidence interval of 95%..

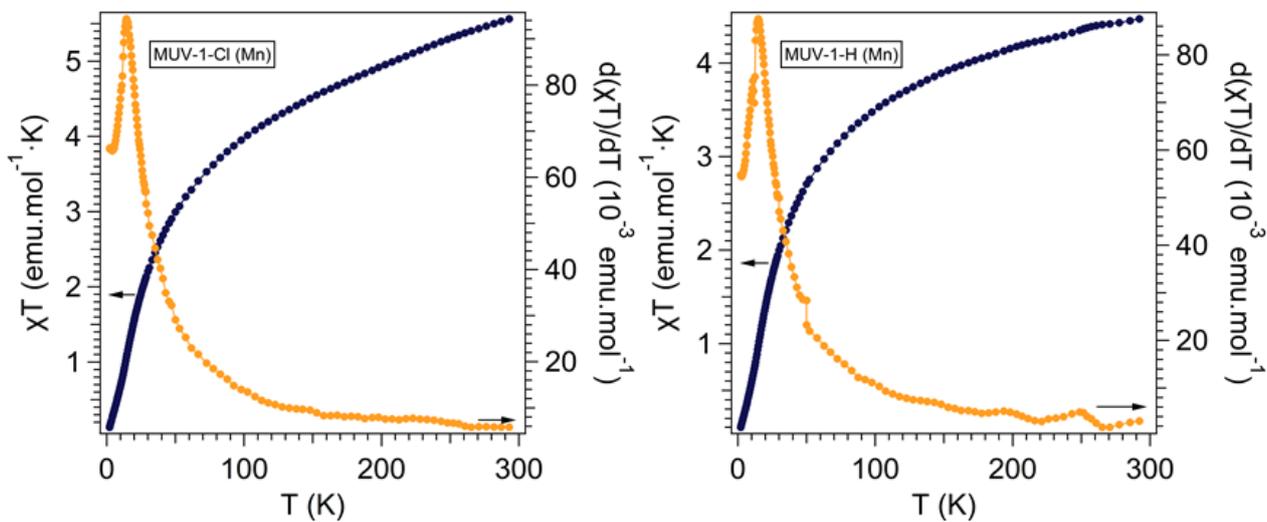

**Figure S15**. Determination of $T_N$ for **MUV-1(Mn)** following the Fisher criteria.[4]

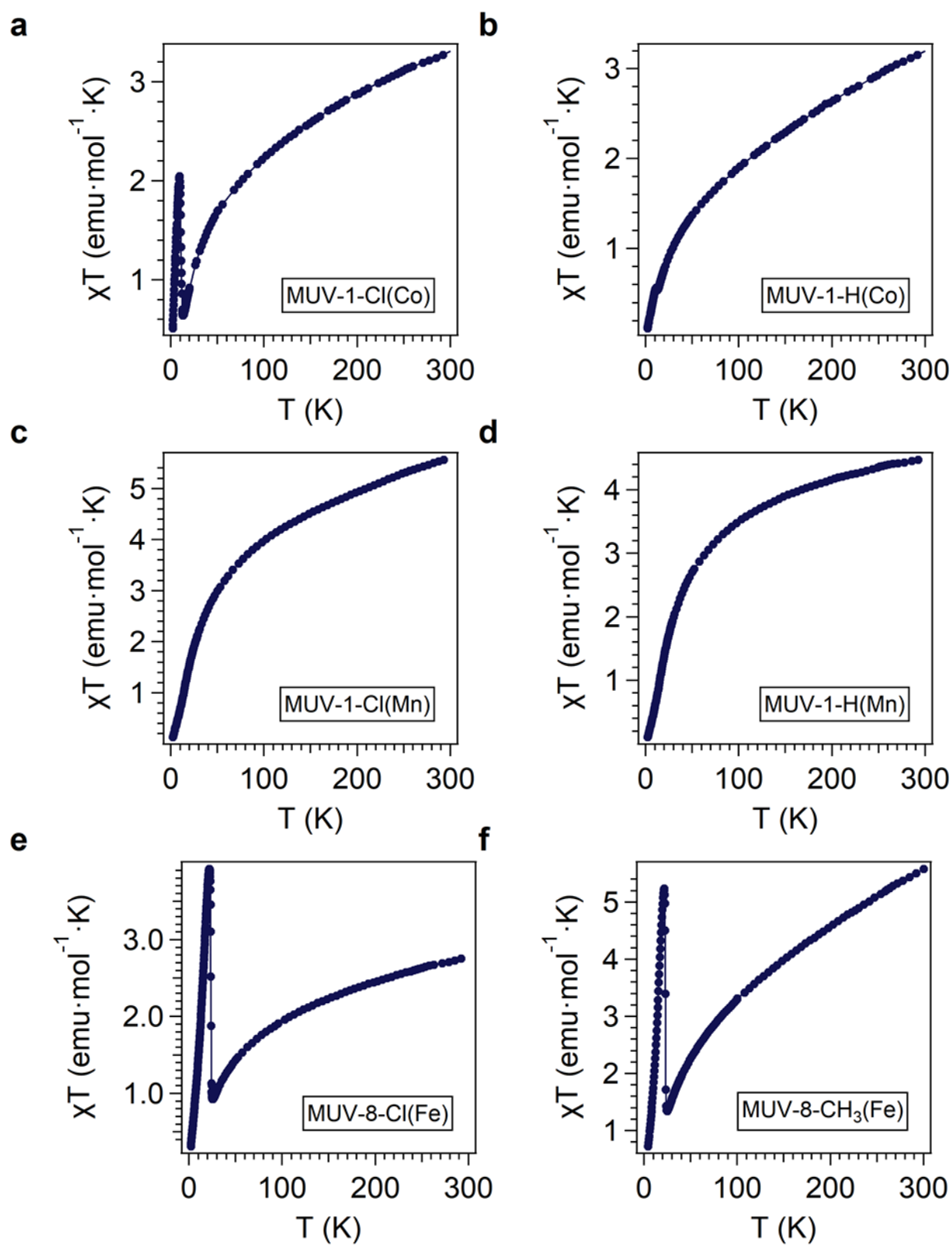

**Figure S16**. Thermal dependence of the product $\chi_m T$ in the temperature range 2-300 K.

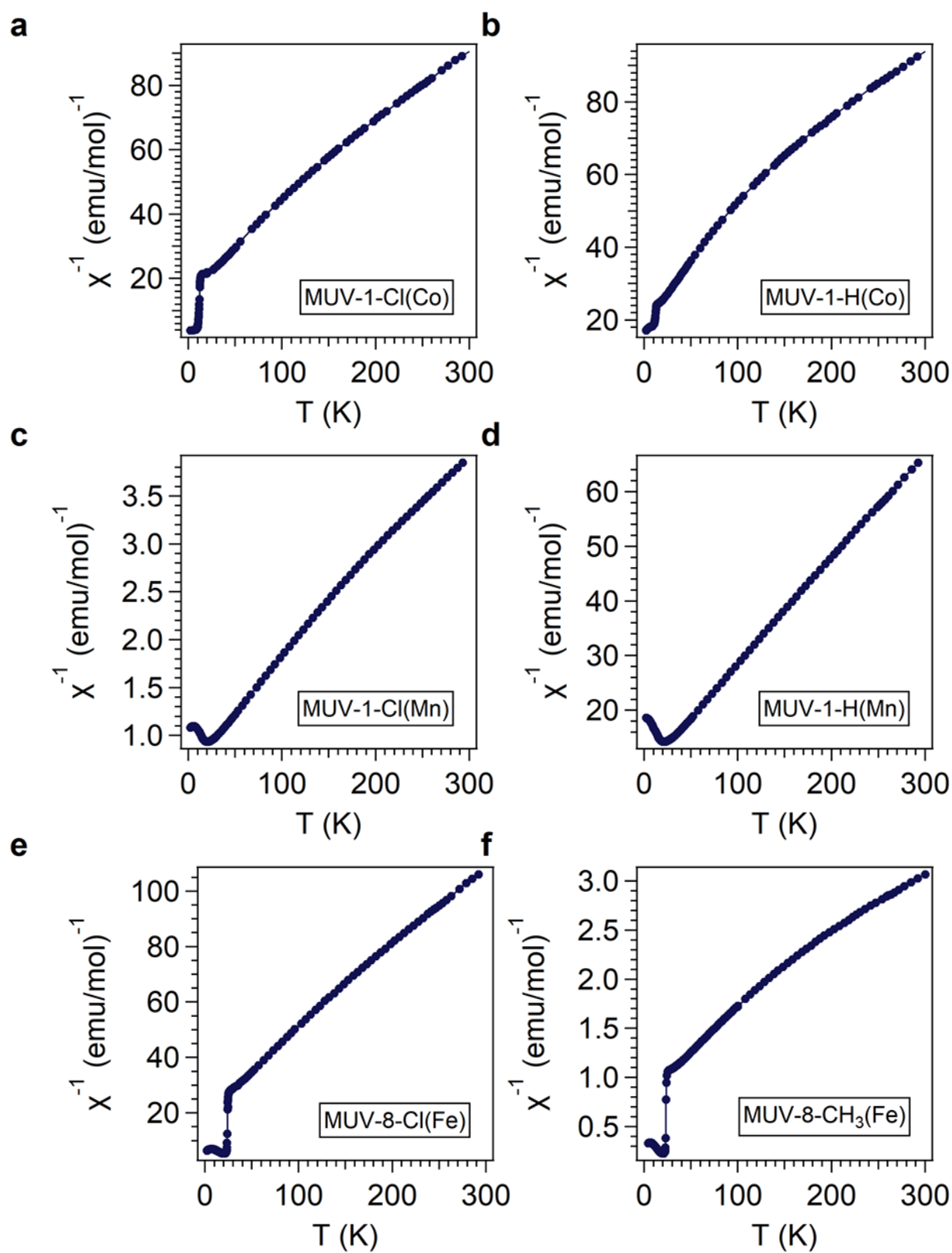

**Figure S17**. Thermal dependence of $\chi_m^{-1}$ in the temperature range 2-300 K.

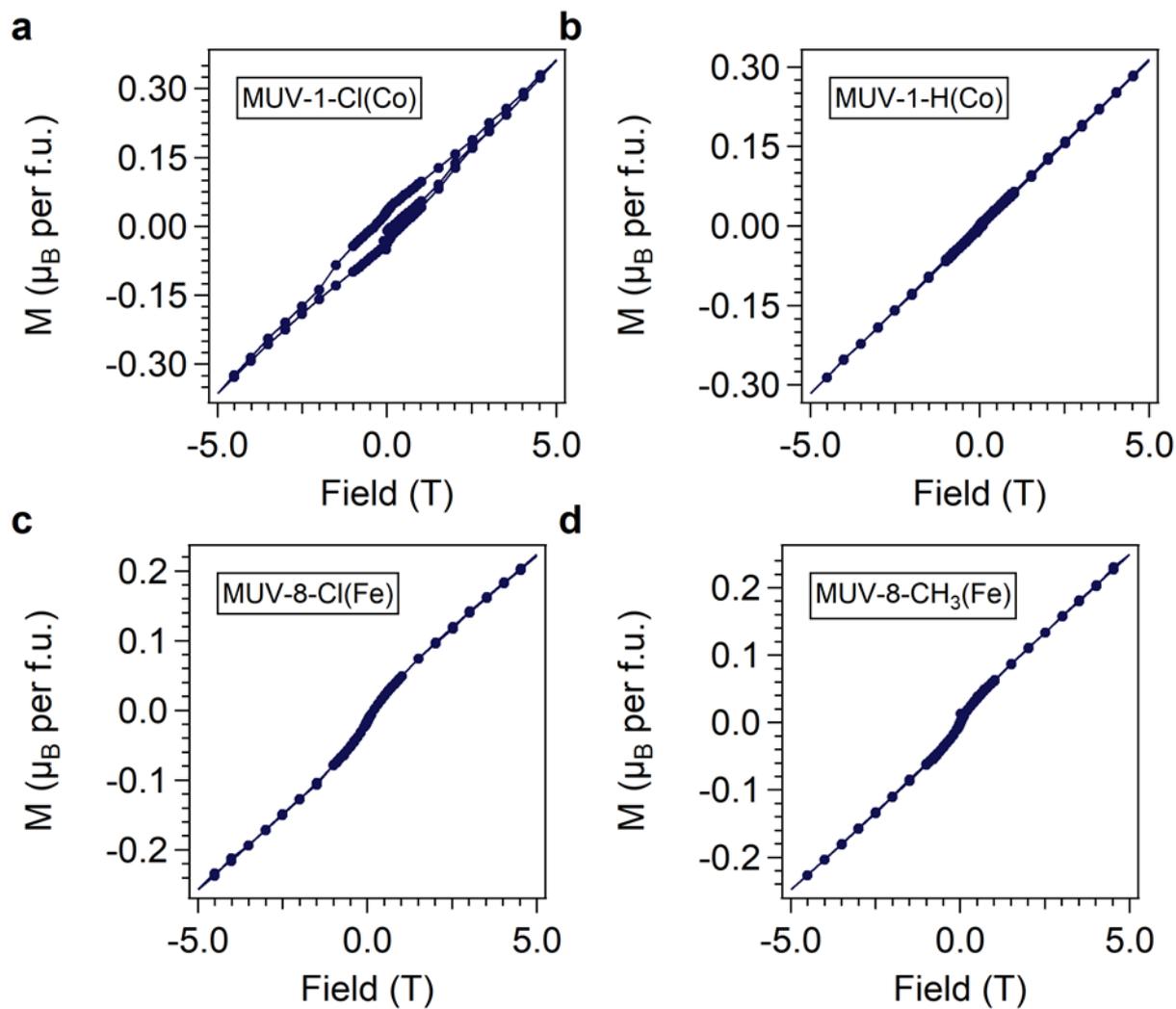

**Figure S18.** Magnetizations at 2 K.

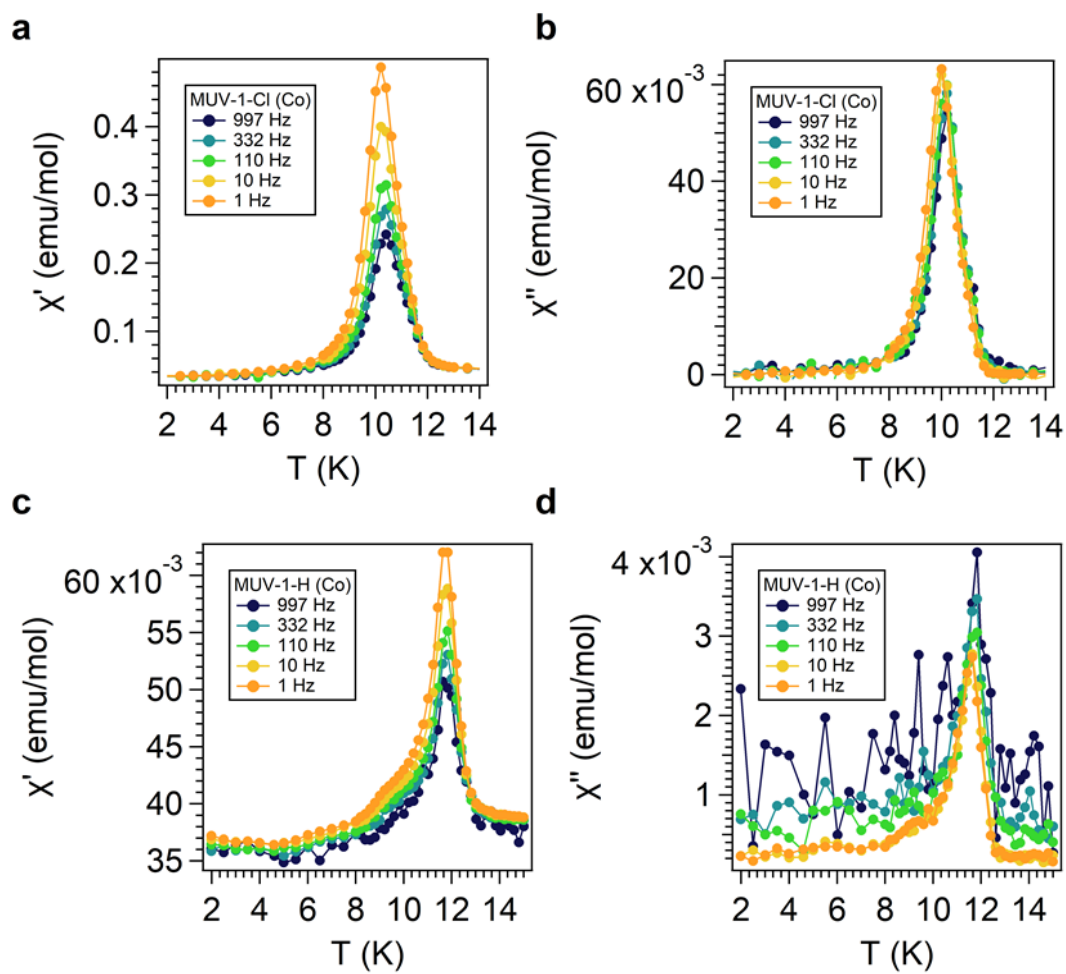

**Figure S19**. In-phase and out-of-phase dynamic susceptibility of **MUV-1(Co)** measured at different frequencies.

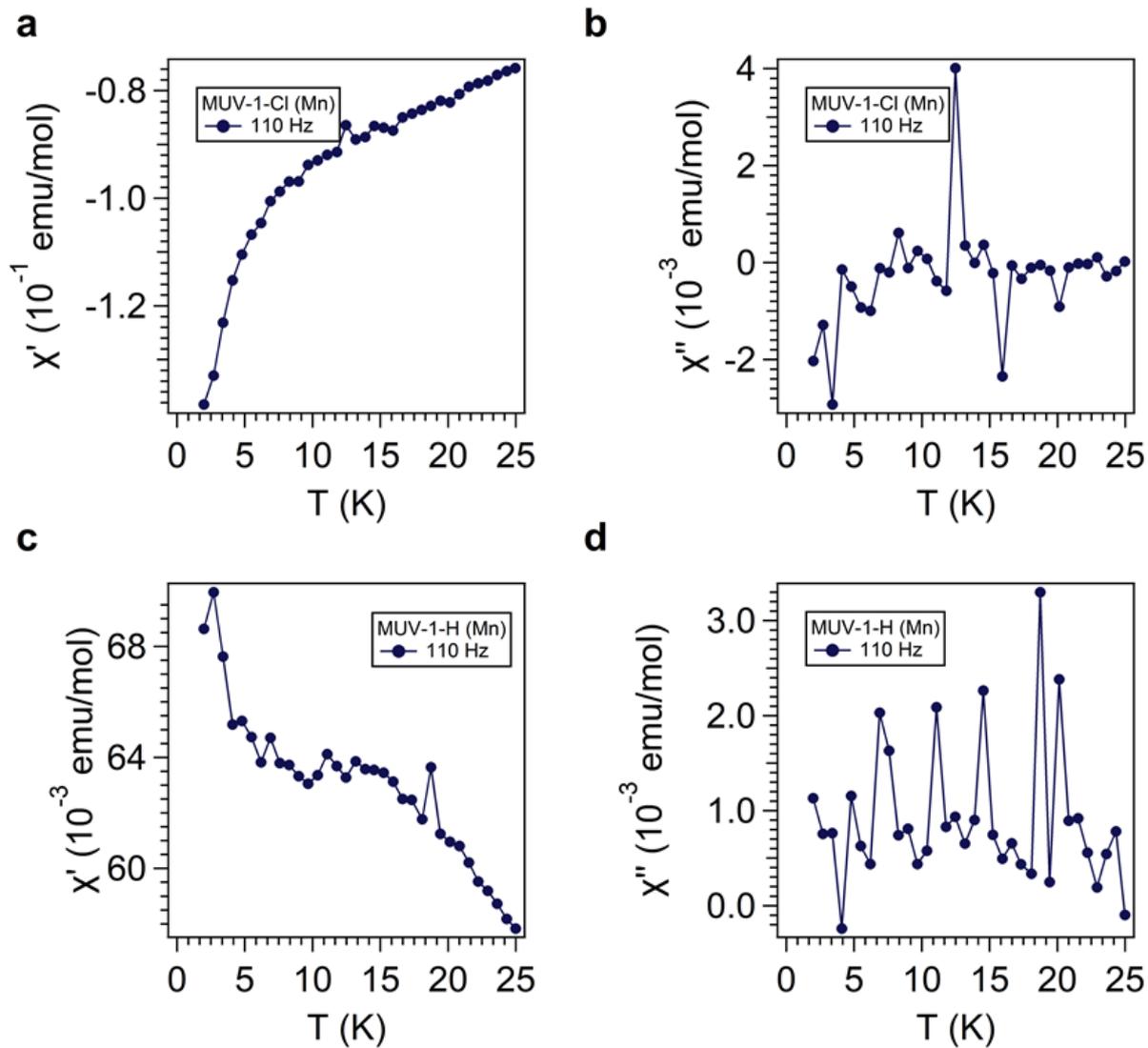

**Figure S20**. In-phase and out-of-phase dynamic susceptibility of **MUV-1(Mn)**.

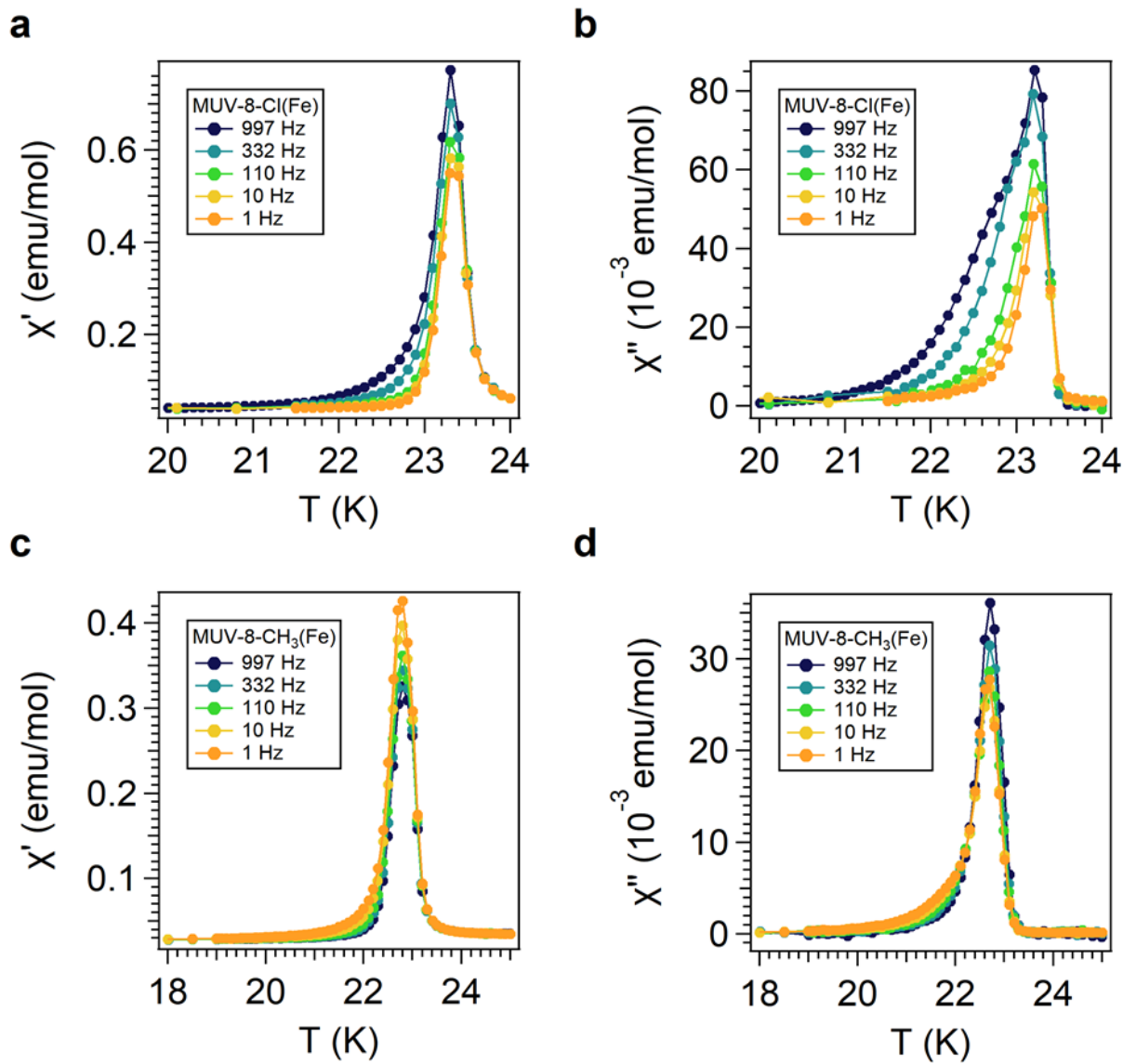

**Figure S21**. In-phase and out-of-phase dynamic susceptibility of **MUV-8** measured at different frequencies.

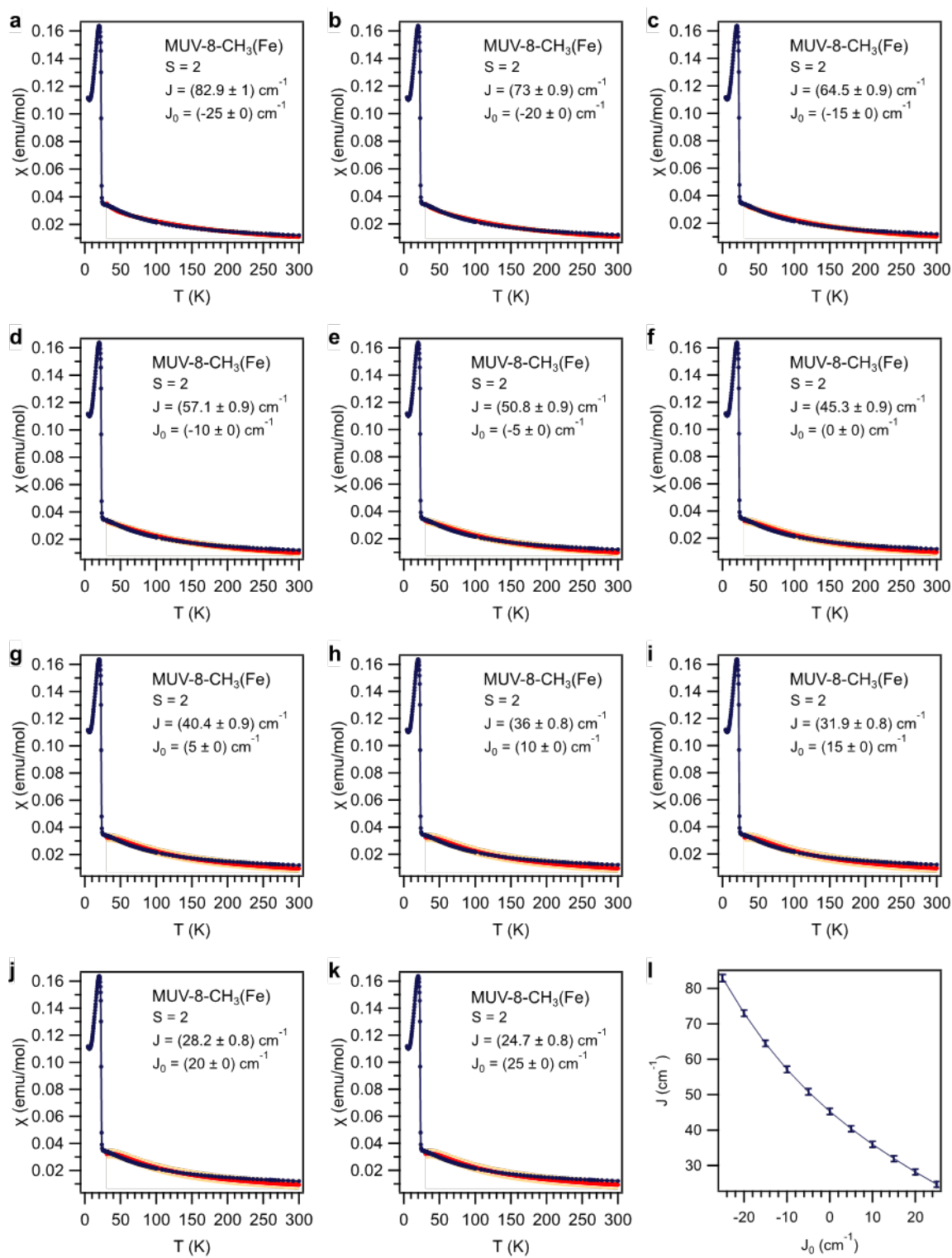

**Figure S22**. J vs $J_0$ phase diagram for **MUV-8-CH$_3$(Fe)**. The data have been fitted following the Curély analysis (solid red line).[6] It is represented in yellowish the prediction bands with a confidence interval of 95%..

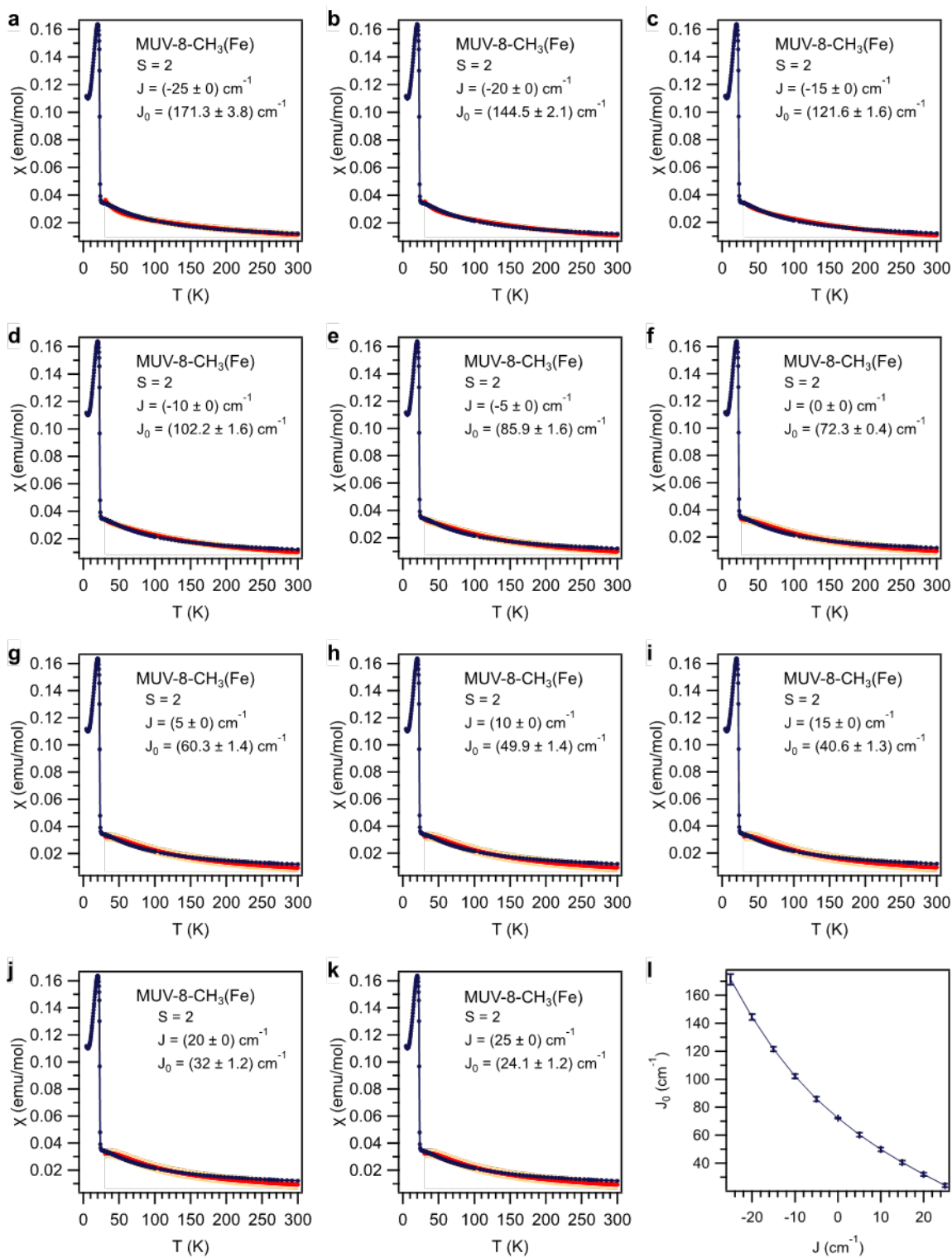

**Figure S23**. $J_0$ vs J phase diagram for **MUV-8-CH₃(Fe)**. The data have been fitted following the Curély analysis (solid red line).[6] It is represented in yellowish the prediction bands with a confidence interval of 95%..

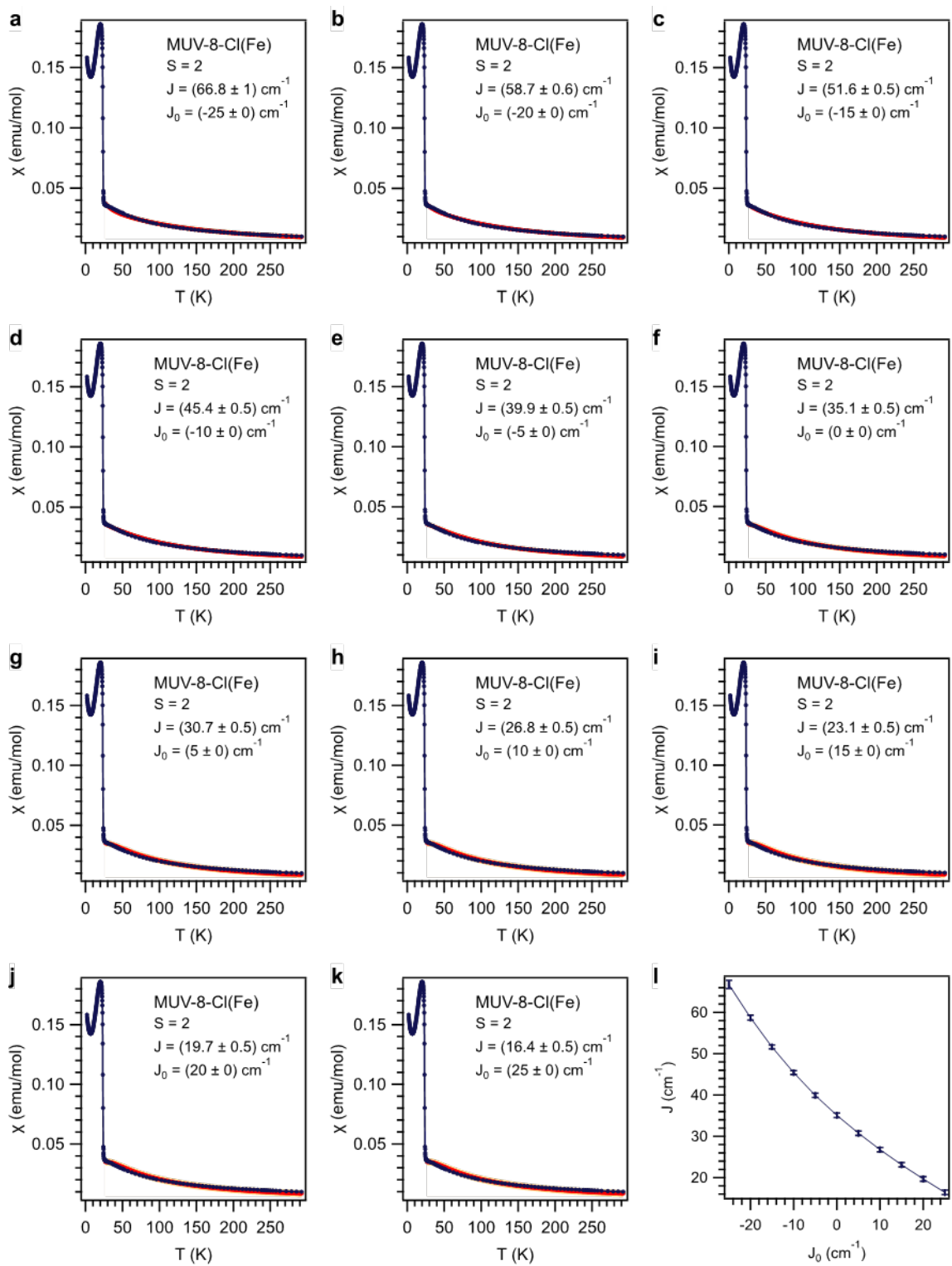

**Figure S24**. J vs $J_0$ phase diagram for **MUV-8-Cl(Fe)**. The data have been fitted following the Curély analysis (solid red line).[6] It is represented in yellowish the prediction bands with a confidence interval of 95%..

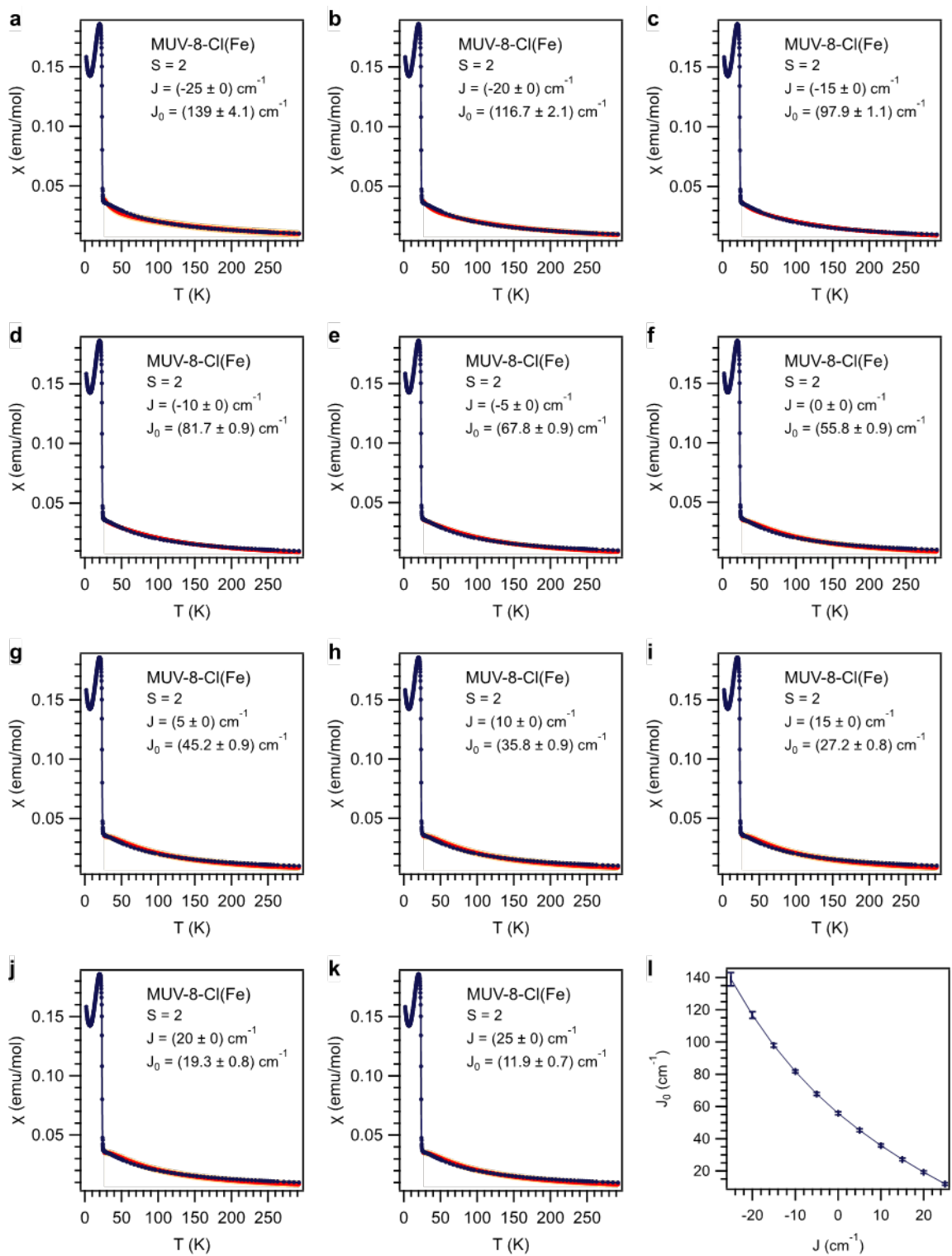

**Figure S25**. $J_0$ vs J phase diagram for **MUV-8-Cl(Fe)**. The data have been fitted following the Curély analysis (solid red line).[6] It is represented in yellowish the prediction bands with a confidence interval of 95%..

**S4.2 Electron Paramagnetic Resonance measurements**

Figure S26 and figure S27 show the EPR spectra of **MUV-1-Cl(Mn)** and **MUV-1-Cl(Co)** for bulk materials at different temperatures. Figure S26 shows a clearly different signal near the ordering temperature (14.3 K). Below this temperature the signal disappears.

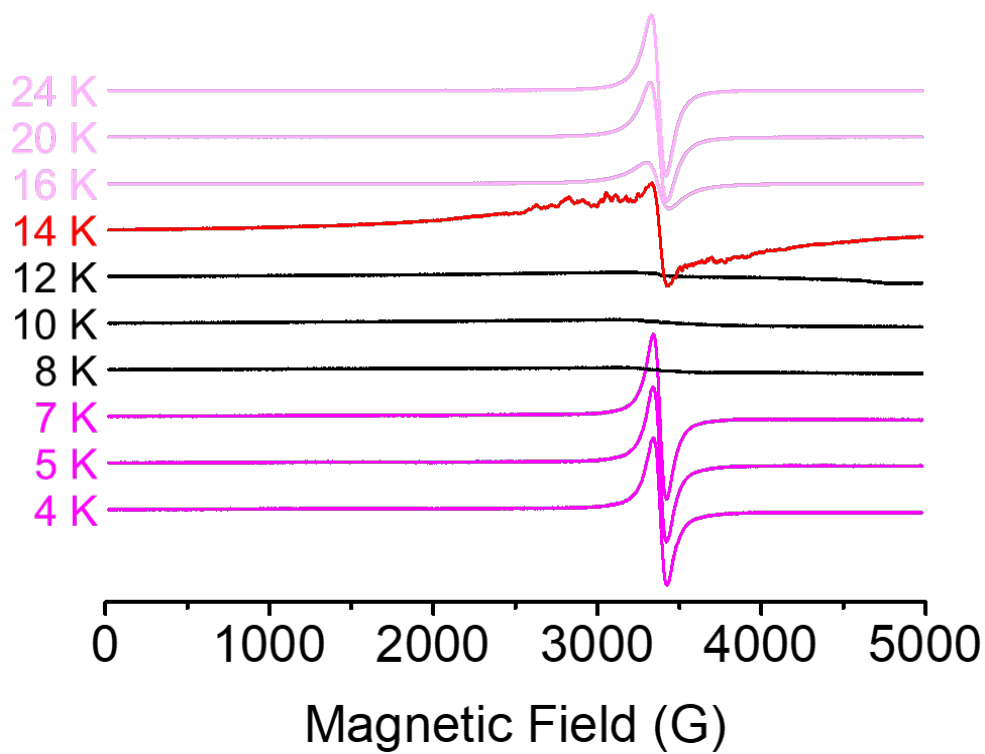

**Figure S26**. EPR spectra of **MUV-1-Cl(Mn)** at different temperatures.

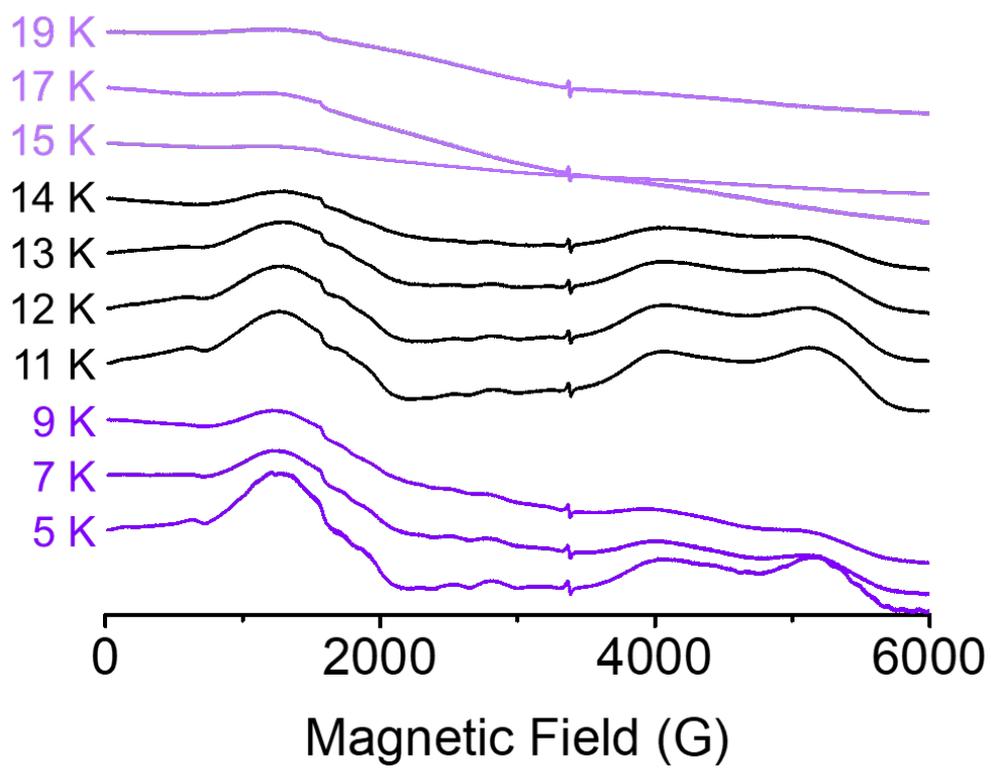

**Figure S27**. EPR spectra **MUV-1-Cl(Co)** at different temperatures.

# S5. Two-dimensional approach and characterization

**MUV-1-Cl(Co), MUV-1-H(Co)**, **MUV-8-Cl(Fe)** and **MUV-8-CH$_3$(Fe)** were mechanically exfoliated by mechanical methods (as commonly known as the *Scotch-tape* method, developed for obtaining graphene from graphite), using an adhesive plastic film of 80 μm thick tape from Ultron Systems. The obtained flakes were deposited onto silicon substrates with 285 nm of thermally grown SiO$_2$ and inspected by different microscopies such us optical microscopy and atomic force microscopy.

## S5.1 Atomic Force Microscopy of the few layers of MUV-1-Cl(Co), MUV-1-H(Co), MUV-8-Cl(Fe) and MUV-8-CH$_3$(Fe)

In this section, it is presented a more detail characterization of the few layers shown in the main manuscript. The thickness of the flakes was determined by atomic force microscopy (Nanoscope IVa Multimode Scanning Probe Microscope, Bruker) in tapping mode.

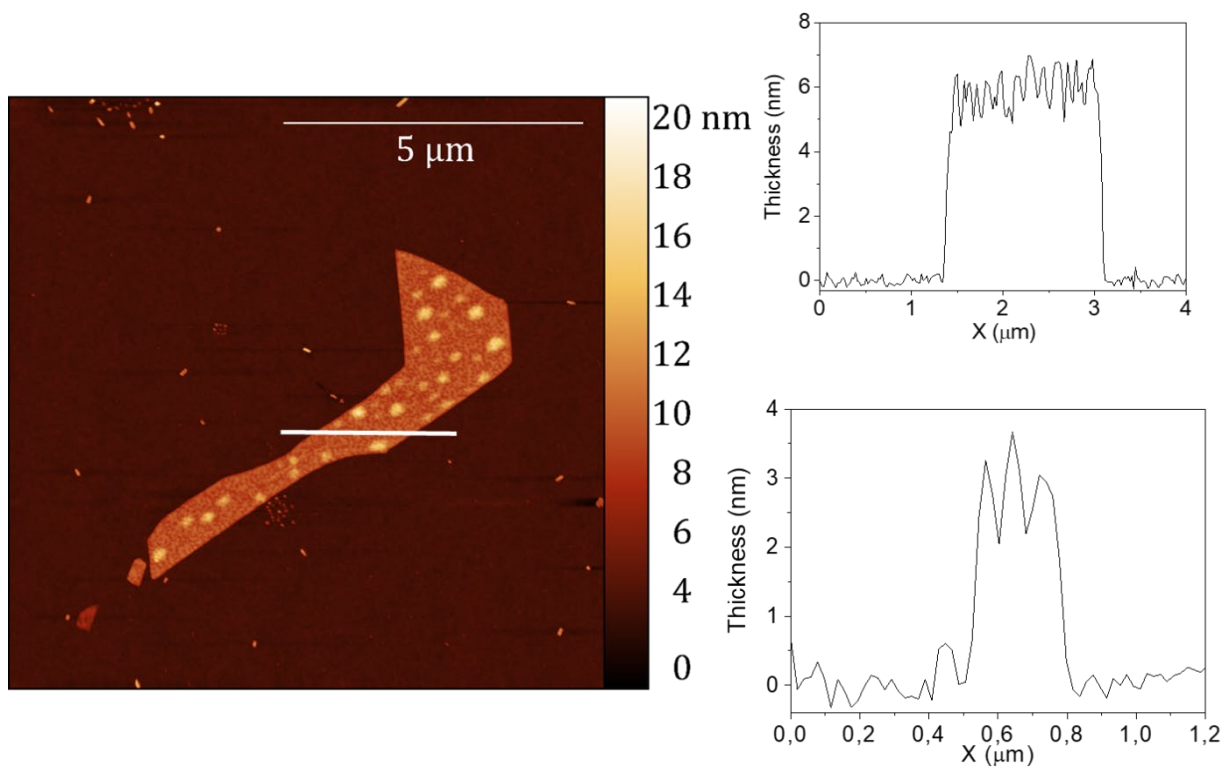

**Figure S28**. Atomic Force Microscopy images of **MUV-1-Cl(Co)** flakes with its corresponding height profiles.

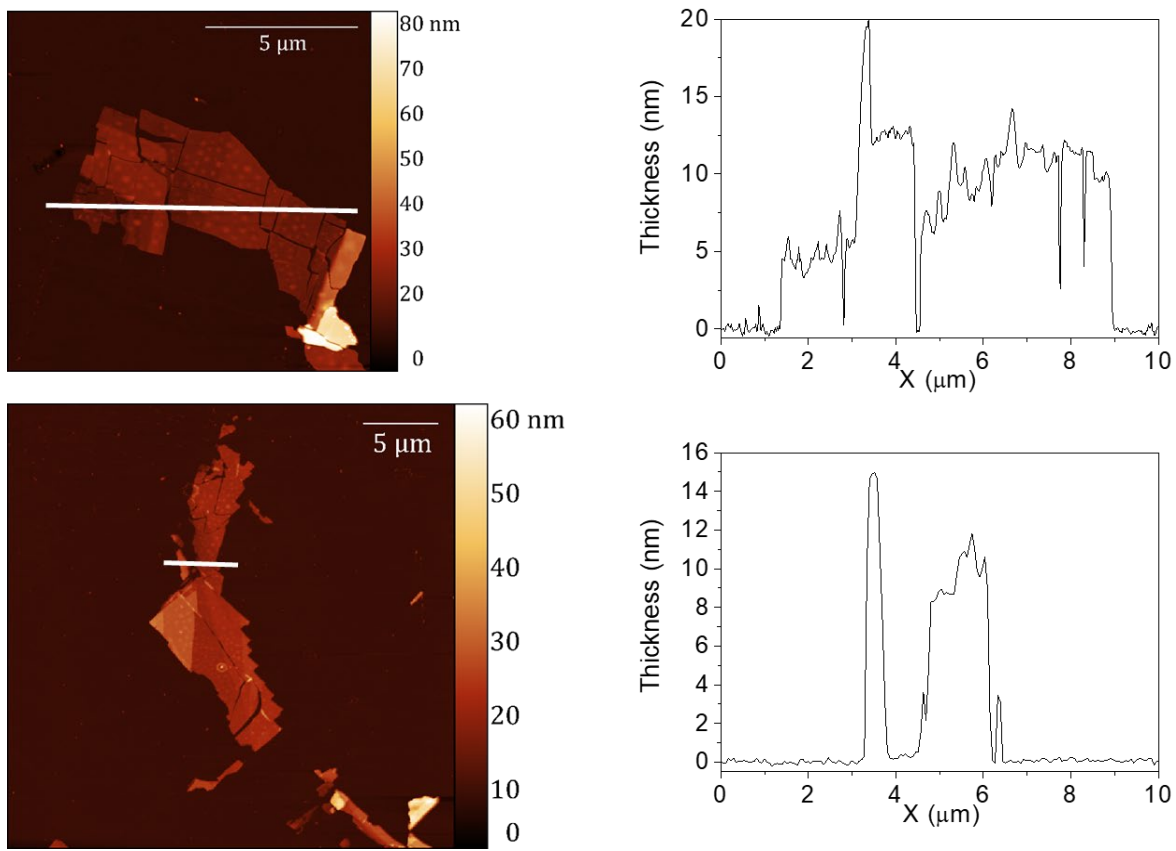

**Figure S29**. Atomic Force Microscopy images of **MUV-1-H(Co)** flakes with its corresponding height profiles.

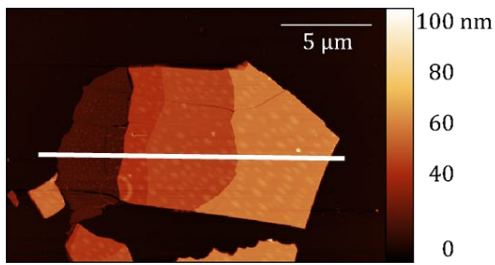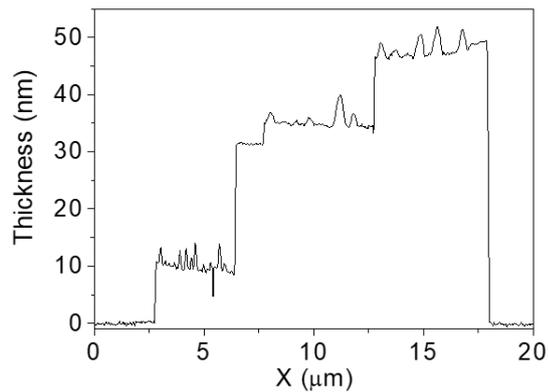
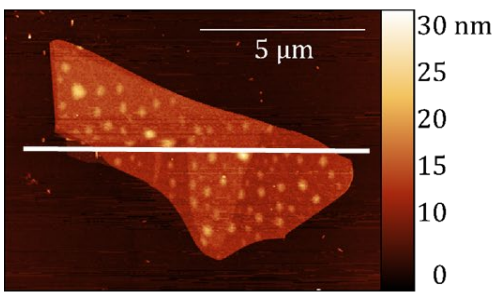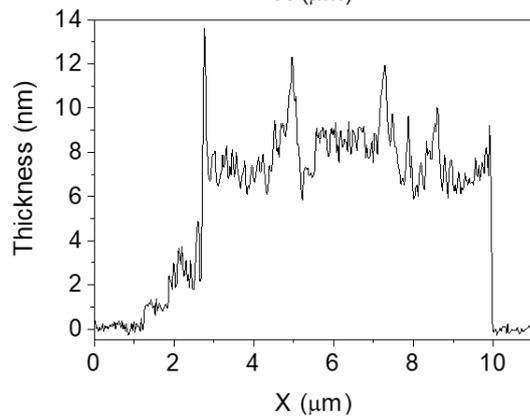

**Figure S30.** Atomic Force Microscopy images of **MUV-8-Cl(Fe)** flakes with its corresponding height profiles.

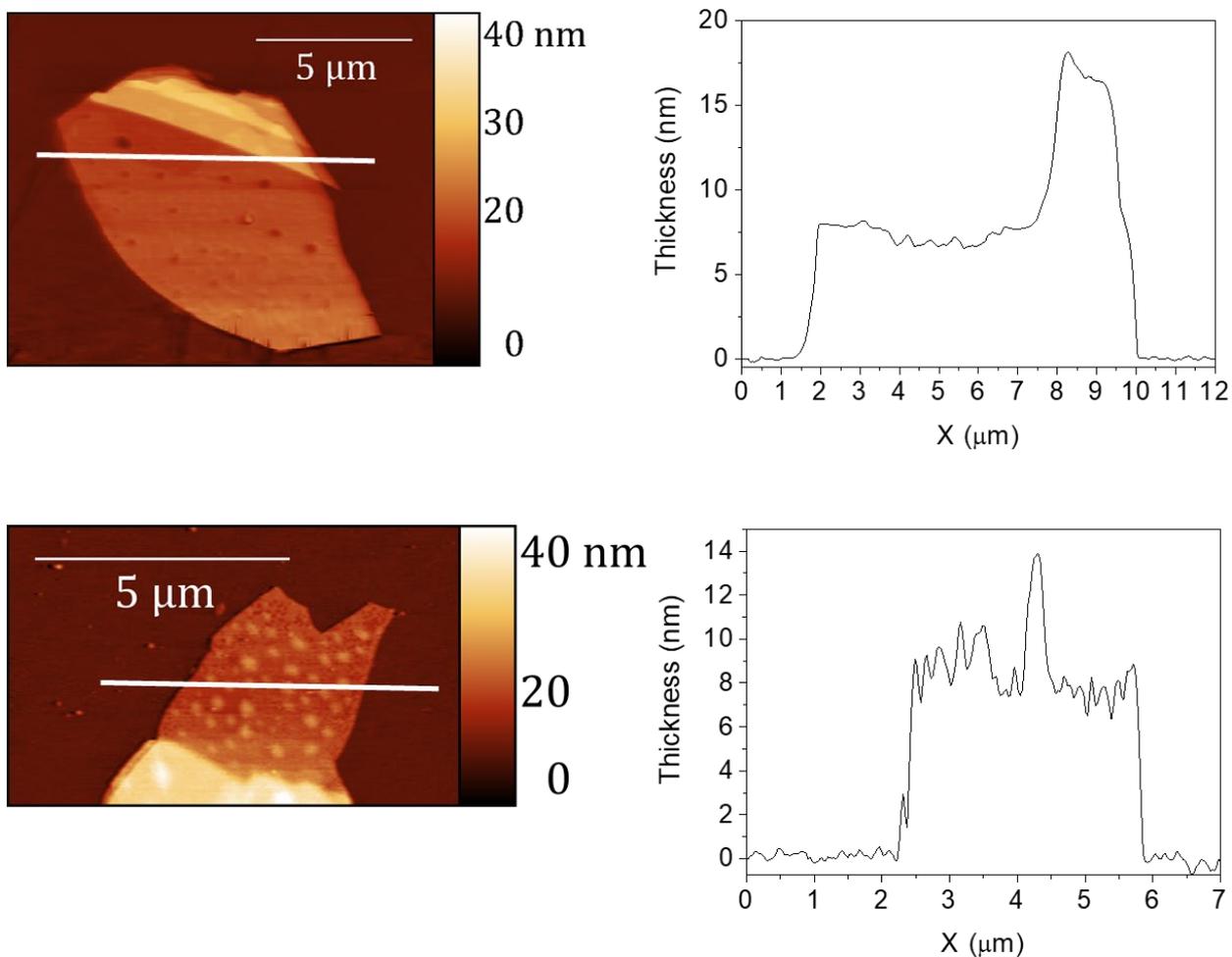

**Figure S31**. Atomic Force Microscopy images of **MUV-8-CH₃(Fe)** flakes with its corresponding height profiles.

**S5.2. Raman spectroscopy in bulk and two-dimensional layers of MUV-1-X(Co) and MUV-8-X(Fe).**

Raman spectra were acquired with a micro-Raman (model XploRA ONE from Horiba, Kyoto, Japan) with a grating of 2400 gr/mm, slit of 50 μm, and hole of 500 μm. The employed wavelengths for the bulk crystals were 532 nm, 638 nm, and 785 nm and 532 nm for the thin-layers. The power density of the laser used for spectra measured at 532 nm was 1.0 mW/μm$^2$, for spectra measured at 638 nm it was and 2.5 mW/μm$^2$ and for those spectra measured at 785 nm it was 3.3 mW/μm$^2$.

The Raman spectrum recorded for the flakes deposited onto Si/285 nm SiO$_2$ substrates were performed employing the wavelength of 532 nm.

In addition, in order to assess the nature of the mechanically exfoliated thin-layers, it was measured the Raman spectra of flakes with different thicknesses on Si/285 nm SiO$_2$ substrates (Figures S3-S34). The thickness of the flakes where the Raman spectra were measured was determined by atomic force microscopy (Nanoscope IVa Multimode Scanning Probe Microscope, Bruker) in tapping mode (Figures S31-S33).

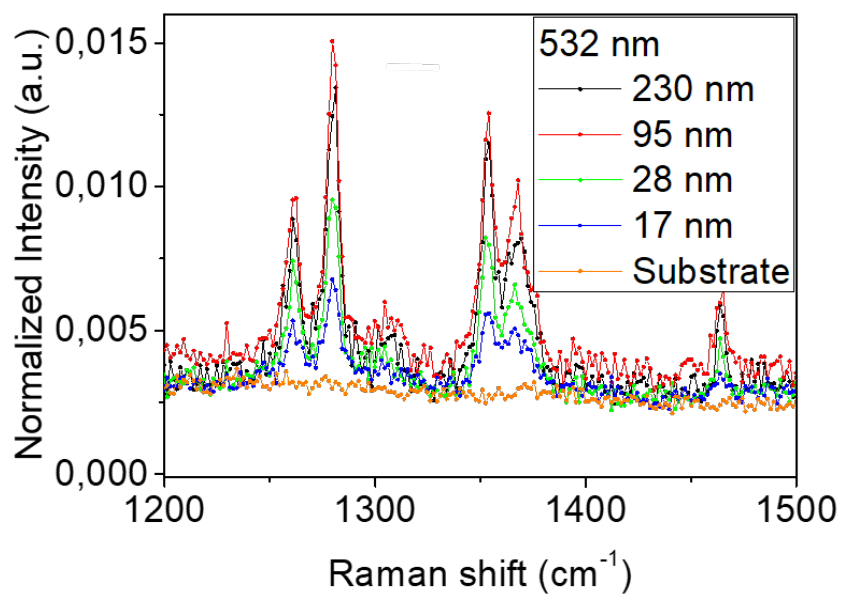

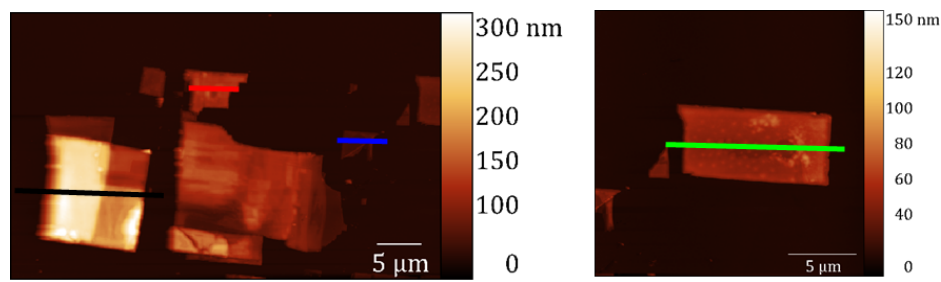

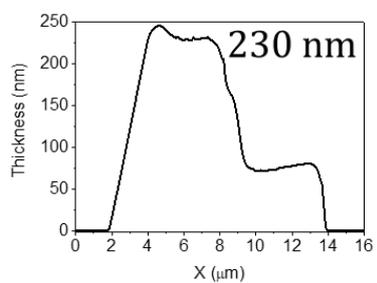
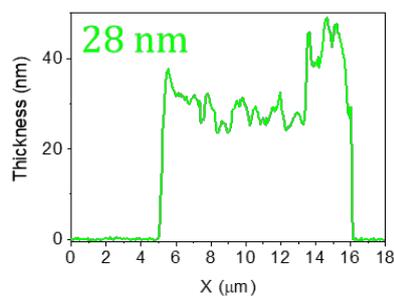
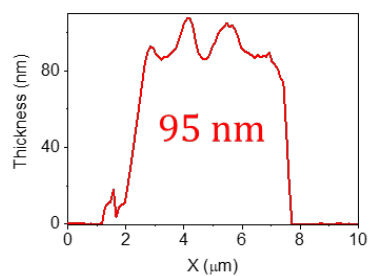
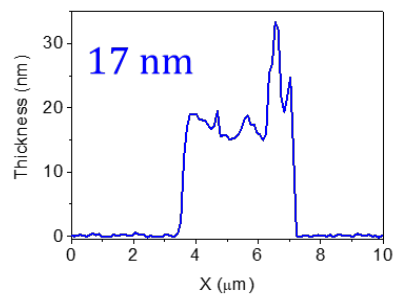

**Figure S32**. Raman spectra and the Atomic Force Microscopy images of **MUV-1-H(Co)** flakes with its corresponding height profiles.

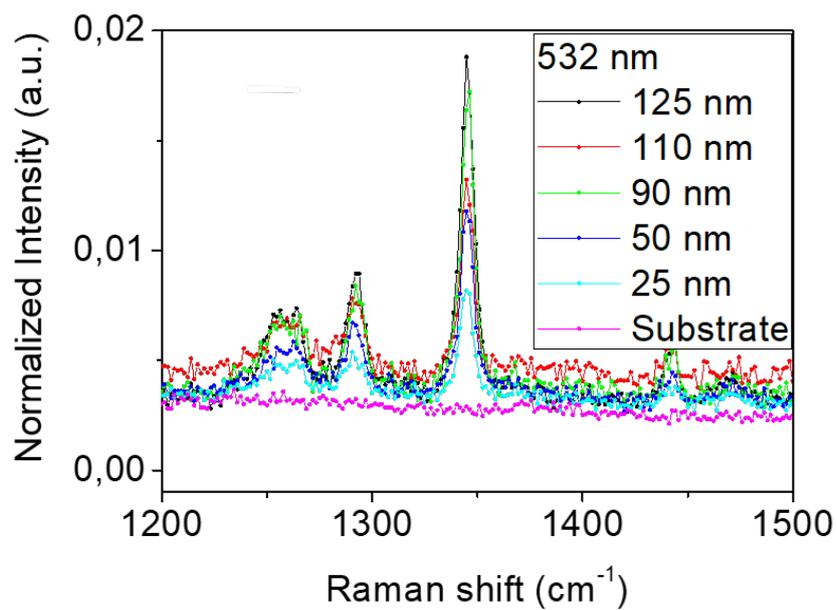

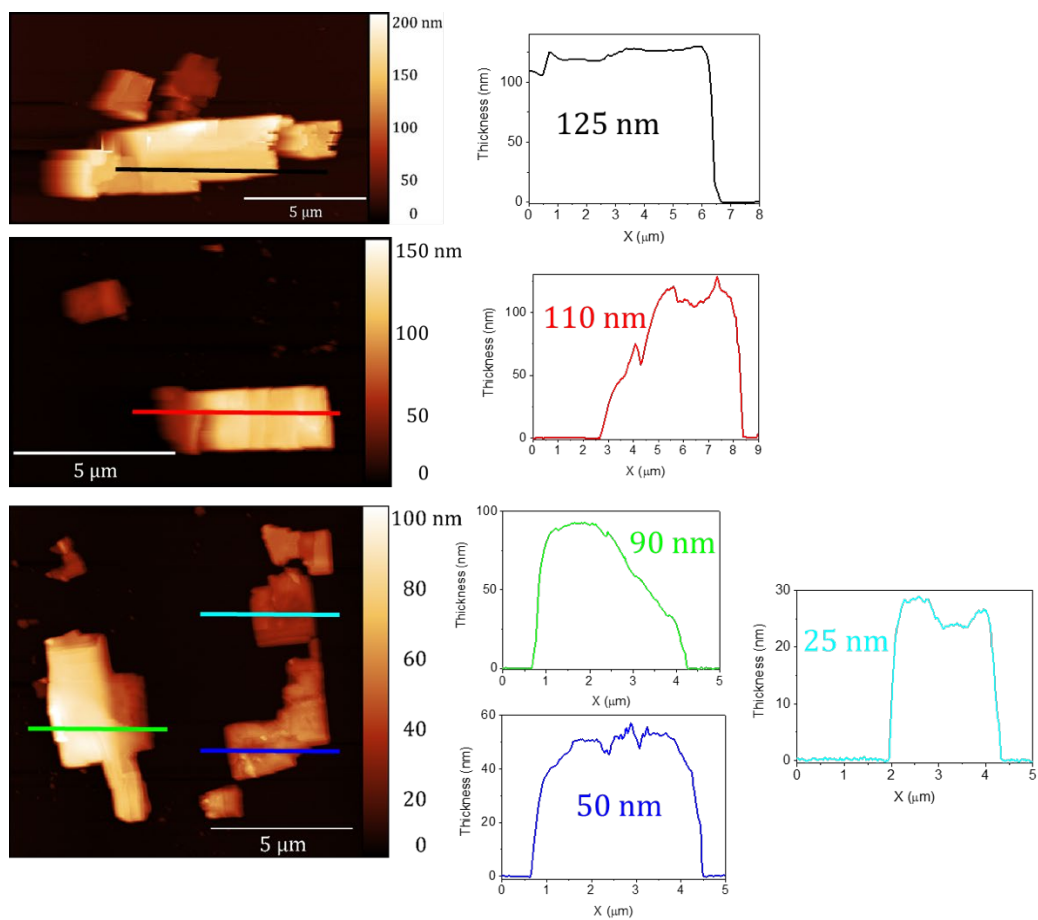

**Figure S33**. Raman spectra and the Atomic Force Microscopy images of **MUV-1-Cl(Co)** flakes with its corresponding height profiles.

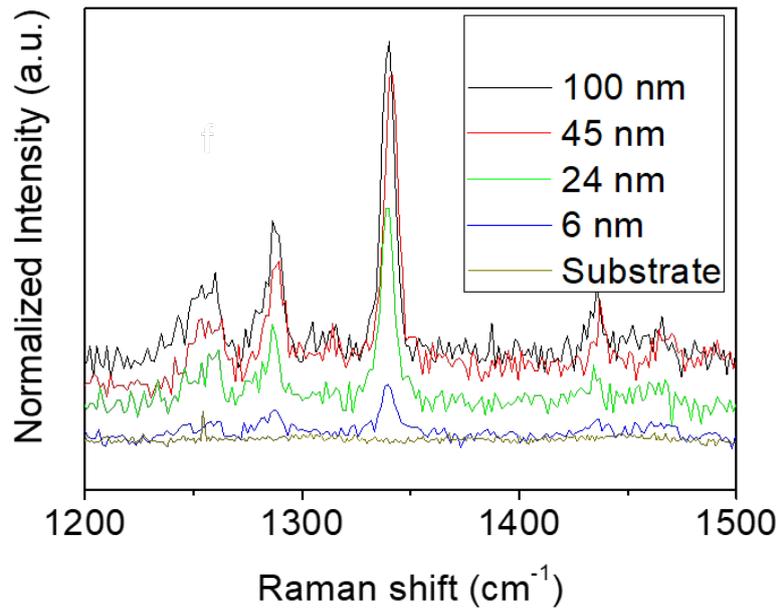

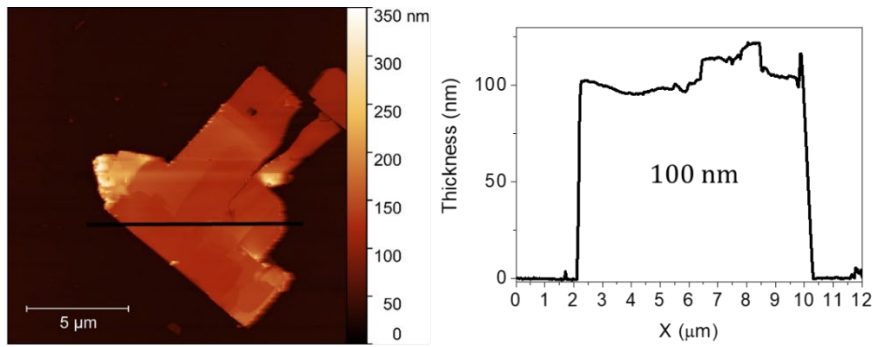

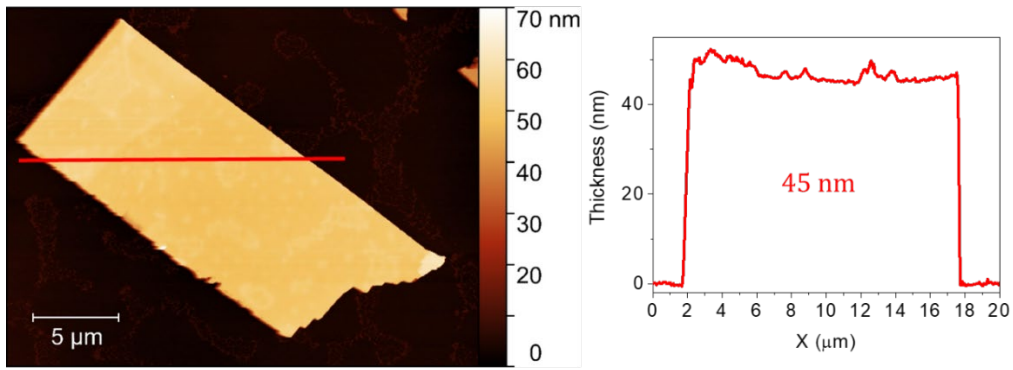

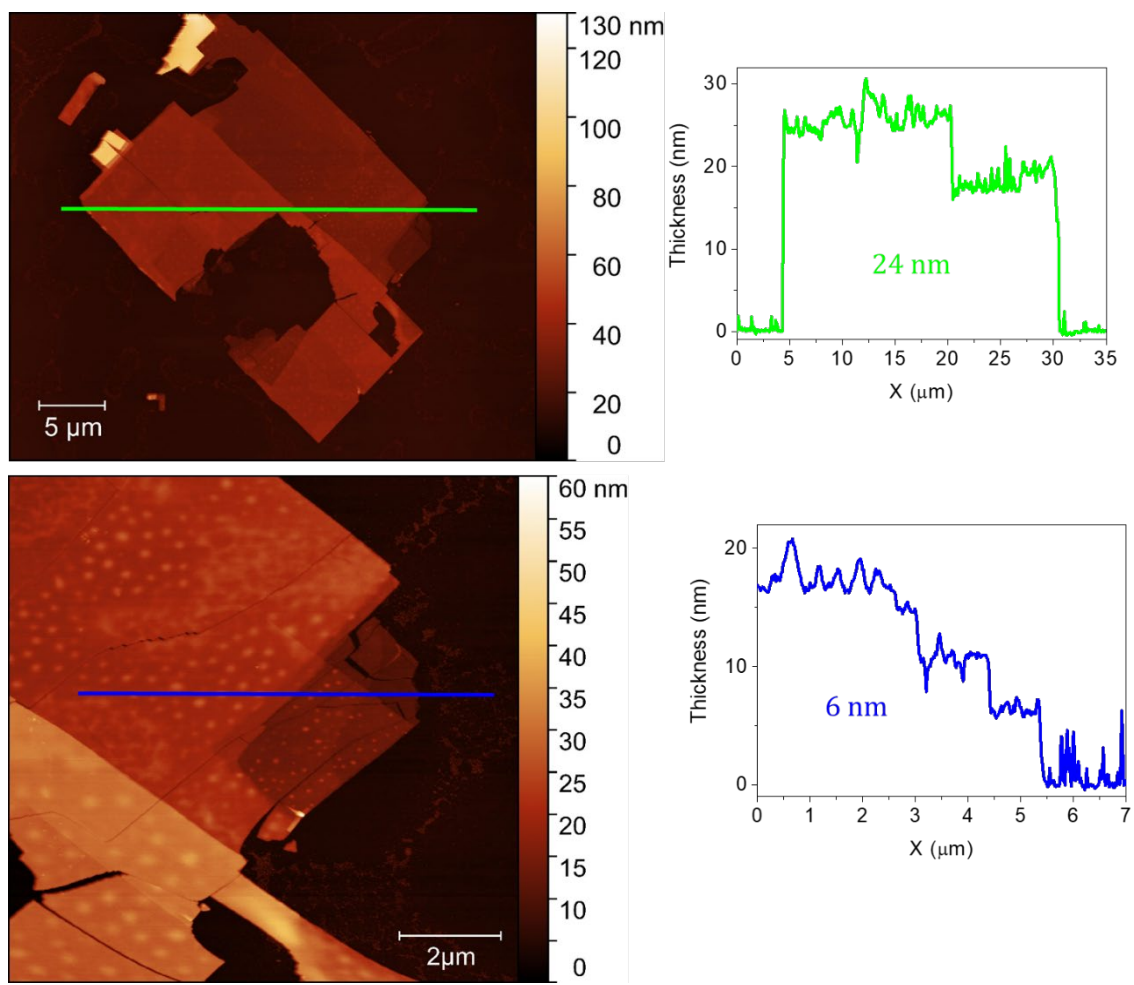

**Figure S34**. Raman spectra and the Atomic Force Microscopy images of **MUV-8-Cl(Fe)** flakes with its corresponding height profiles.